\documentstyle[amssymb,prl,aps,epsfig]{revtex}

\begin{document}

\title{Relationship between and implications of the isotope and pressure effects on transition
temperature, penetration depths and conductivities. }
\author{T. Schneider}
\address{Physik-Institut der Universit\"{a}t Z\"{u}rich, Winterthurerstr. 190, CH-8057%
\\
Z\"urich, Switzerland}
\maketitle

\begin{abstract}
It is shown that the empirical relations between transition
temperature, normal state conductivity linearly extrapolated to
the value at the transition temperature, zero temperature
penetration depths, etc., as observed in a rich variety of cuprate
superconductors, are remarkably consistent with the universal
critical properties of anisotropic systems which fall into the
3D-XY universality class and undergo a crossover to a quantum
critical point in 2D. The variety includes n- and p-type cuprates,
comprises the underdoped and overdoped regimes and the consistency
extends up to six decades in the scaling variables. The resulting
scaling relations for the oxygen isotope hydrostatic pressure
effects agree with the experimental data and reveal that these
effects originate from local lattice distortions preserving the
volume of the unit cell. These observations single out 3D and
anisotropic microscopic models which incorporate local lattice
distortions, fall in the experimentally accessible regime into the
3D-XY universality class, and incorporate the crossover to 2D
quantum criticality where superconductivity disappears.
\end{abstract}

\bigskip

\section{Introduction}

\begin{figure}[tbp]
\centering
\includegraphics[totalheight=6cm]{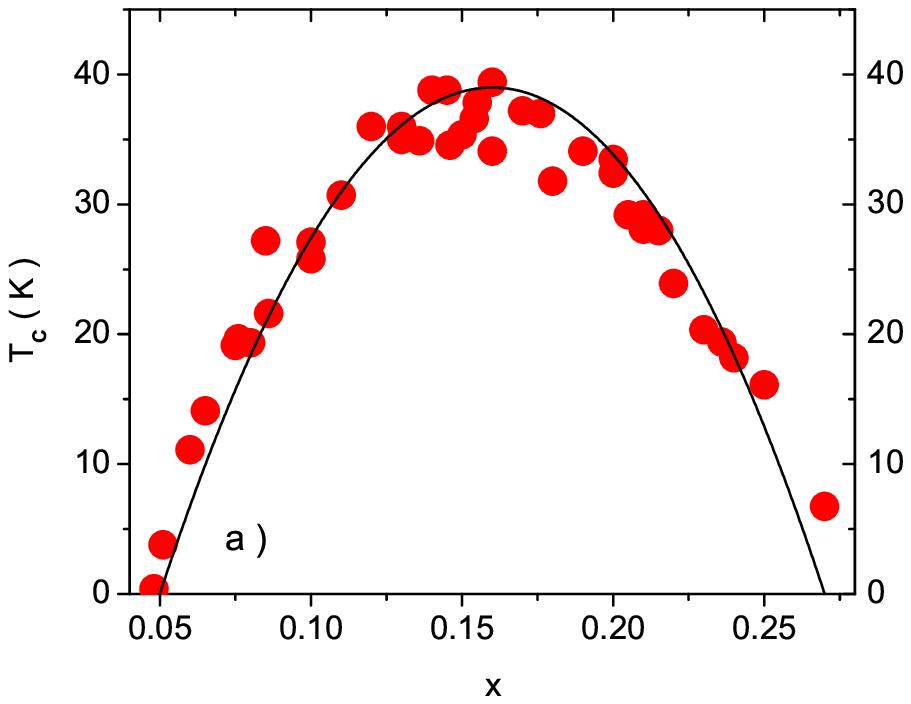}
\includegraphics[totalheight=6cm]{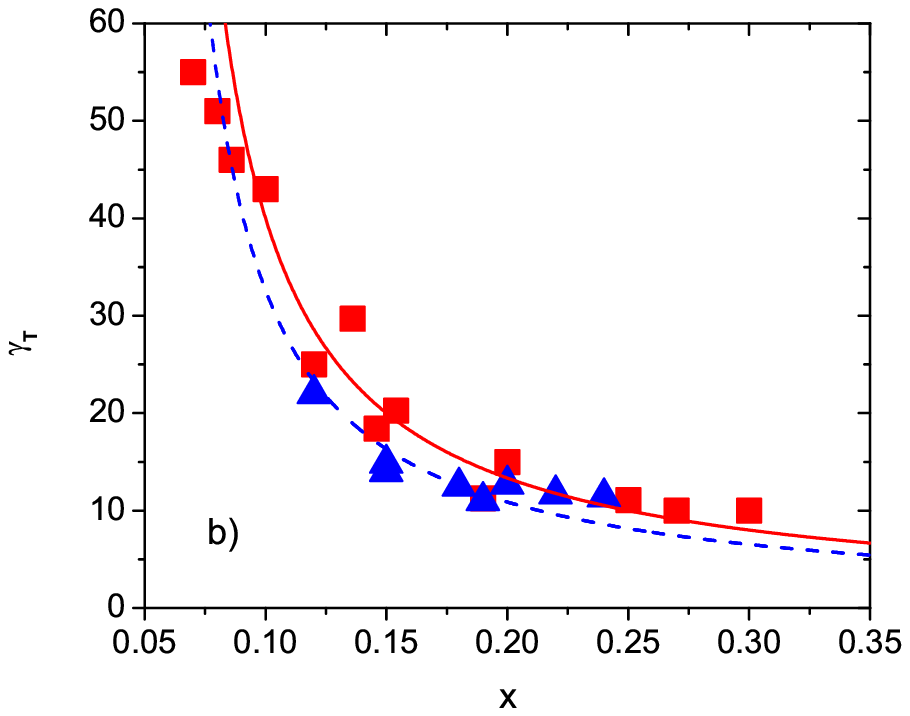}
\caption{a) Variation of $T_{c}$ for La$_{2-x}$Sr$_{x}$CuO$_{4}$.
Experimental data taken from \protect\cite
{suzuki,nakamura,fukuzumi,willemin,kimura,sasagawa,hoferdis,shibauchi,panagopoulos}.
The solid line is Eq.(\ref{eq1}) with $T_{c}\left( x_{m}\right)
=39$K. b) $\gamma _{T}$ versus $x$ for
La$_{2-x}$Sr$_{x}$CuO$_{4}$. The squares are the experimental data
for $\gamma _{T_{c}}$ \protect\cite
{suzuki,nakamura,willemin,sasagawa,hoferdis} and the triangles for
$\gamma _{T=0}$ \protect\cite{shibauchi,panagopoulos}. The solid
curve and dashed lines are Eq. (\ref{eq2}) with $\gamma
_{T_{c},0}=2$ and $\gamma _{T=0,0}=1.63$.} \label{fig1}
\end{figure}

Establishing and understanding the phase diagram of cuprate
superconductors in the temperature - dopant concentration plane is
one of the major challenges in condensed matter physics.
Superconductivity is derived from the insulating and
antiferromagnetic parent compounds by partial substitution of ions
or by adding or removing oxygen. For instance La$_{2}$CuO$_{4}$
can be doped either by alkaline earth ions or oxygen to exhibit
superconductivity. The empirical phase diagram of
La$_{2-x}$Sr$_{x}$CuO$_{4}$ \cite
{suzuki,nakamura,fukuzumi,willemin,kimura,sasagawa,hoferdis,shibauchi,panagopoulos}
depicted in Fig. \ref{fig1} shows that after passing the so called
underdoped limit $\left( x_{u}\approx 0.047\right) $, $T_{c}$
reaches its maximum value $T_{c}^{m}$ at $x_{m}\approx 0.16$. With
further increase of $x $, $T_{c}$ decreases and finally vanishes
in the overdoped limit $x_{o}\approx 0.273$.

This phase transition line is thought to be a generic property of
cuprate superconductors \cite{tallon} \ and is well described by
the empirical relation
\begin{equation}
T_{c}\left( x\right) =T_{c}\left( x_{m}\right) \left( 1-2\left(
\frac{x}{x_{m}}-1\right) ^{2}\right) =\frac{2T_{c}\left(
x_{m}\right) }{x_{m}^{2}}\left( x-x_{u}\right) \left(
x_{o}-x\right) ,\ \ x_{m}=0.16, \label{eq1}
\end{equation}
proposed by Presland \emph{et al}.\cite{presland}. Approaching the
endpoints along the axis $x$, La$_{2-x}$Sr$_{x}$CuO$_{4}$
undergoes at zero temperature doping tuned quantum phase
transitions. As their nature is concerned, resistivity
measurements\cite{fukuzumi,momono} reveal a quantum superconductor
to insulator (QSI) transition in the underdoped limit\cite
{polen,book,klosters,parks} and in the overdoped limit a quantum
superconductor to normal state (QSN) transition\cite
{polen,book,klosters,parks}.

Another essential experimental fact is the doping dependence of
the anisotropy. In tetragonal cuprates it is defined as the ratio
$\gamma =\xi _{ab}/\xi _{c}$ of the correlation lengths parallel
$\left( \xi _{ab}\right) $ and perpendicular $\left( \xi
_{c}\right) $ to CuO$_{2}$ layers (ab-planes). In the
superconducting state it can also be expressed as the ratio
$\gamma =\lambda _{c}/\lambda _{ab}$ of the London penetration
depths due to supercurrents flowing perpendicular ($\lambda _{c}$
) and parallel ($\lambda _{ab}$ ) to the ab-planes. Approaching a
non-superconductor to superconductor transition $\xi $ diverges,
while in a superconductor to non-superconductor transition
$\lambda $ tends to infinity. In both cases, however, $\gamma $
remains finite as long as the system exhibits anisotropic but
genuine 3D behavior. There are two limiting cases: $\gamma =1$
characterizes isotropic 3D- and $\gamma =\infty $ 2D-critical
behavior. An instructive model where $\gamma $ can be varied
continuously is the anisotropic 2D Ising model\cite{onsager}. When
the coupling in the y direction goes to zero, $\gamma =\xi
_{x}/\xi _{y}$ becomes infinite, the model reduces to the 1D case
and $T_{c}$ vanishes. In the Ginzburg-Landau description of
layered superconductors the anisotropy is related to the
interlayer coupling. The weaker this coupling is, the larger
$\gamma $ is. The limit $\gamma =\infty $ is attained when the
bulk superconductor corresponds to a stack of independent slabs of
thickness $d_{s}$. With respect to experimental work, a
considerable amount of data is available on the chemical
composition dependence of $\gamma $ . At $T_{c}$ it can be
inferred from resistivity ($\gamma =\xi _{ab}/\xi _{c}=\sqrt{\rho
_{ab}/\rho _{c}}$) and magnetic torque measurements, while in the
superconducting state it follows from magnetic torque and
penetration depth ($\gamma =\lambda _{c}/\lambda _{ab}$) data. In
Fig. \ref{fig1}b we displayed the doping dependence of $1/\gamma
_{T}$ evaluated at $T_{c}$ ($\gamma _{T_{c}}$) and $T=0$ ($\gamma
_{T=0}$). As the dopant concentration is reduced, $\gamma
_{T_{c}}$ and $\gamma _{T=0}$ increase systematically, and tend to
diverge in the underdoped limit. Thus the temperature range where
superconductivity occurs shrinks in the underdoped regime with
increasing anisotropy. This competition between anisotropy and
superconductivity raises serious doubts whether 2D mechanisms and
models\cite{anderson}, corresponding to the limit $\gamma
_{T}=\infty $, can explain the essential observations of
superconductivity in the cuprates. From Fig. \ref{fig1}b it is
also seen that $\gamma _{T}\left( x\right) $ is well described
by\cite{parks,tsphysb}
\begin{equation}
\gamma _{T}\left( x\right) =\frac{\gamma _{T,0}}{x-x_{u}}.
\label{eq2}
\end{equation}
Having also other cuprate families in mind, it is convenient to
express the dopant concentration in terms of $T_{c}$. From Eqs.
(\ref{eq1}) and(\ref{eq2}) we obtain the correlation between
$T_{c}$ and $\gamma _{T}$:
\begin{equation}
\frac{T_{c}}{T_{c}\left( x_{m}\right) }=1-\left( \frac{\gamma
_{T}\left( x_{m}\right) }{\gamma _{T}}-1\right) ^{2},\ \ \gamma
_{T}\left( x_{m}\right) =\frac{\gamma _{T,0}}{x_{m}-x_{u}}
\label{eq3}
\end{equation}

\begin{figure}[tbp]
\centering
\includegraphics[totalheight=6cm]{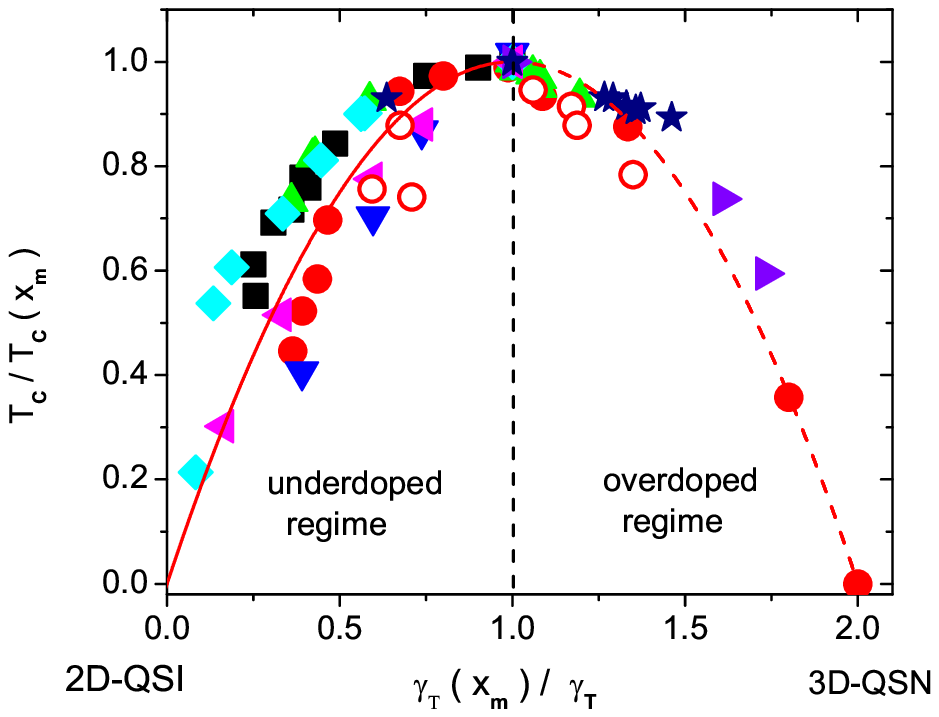}
\caption{$\ \ T_{c}/T_{c}\left( x_{m}\right) $ versus $\gamma
_{T}\left( x_{m}\right) /$ $\gamma _{T}$ for
La$_{2-x}$Sr$_{x}$CuO$_{4}$ ($\bullet $, $T_{c}\left( x_{m}\right)
=37$K, $\gamma _{T_{c}}\left( x_{m}\right) =20$)
\protect\cite{suzuki,nakamura,willemin,sasagawa,hoferdis}\ ,
($\bigcirc $, $T_{c}\left( x_{m}\right) =37$K, $\gamma
_{T=0}\left( x_{m}\right) =14.9$)
\protect\cite{shibauchi,panagopoulos}, HgBa$_{2}$CuO$_{4+\delta }$
($\blacktriangle $, $T_{c}\left( x_{m}\right) =95.6$K, $\gamma
_{T_{c}}\left( x_{m}\right) =27$) \protect\cite{hoferhg},
Bi$_{2}$Sr$_{2}$CaCu$_{2}$O$_{8+\delta }$ ($\bigstar $,
$T_{c}\left( x_{m}\right) =84.2$K, $\gamma _{T_{c}}\left(
x_{m}\right) =133$) \protect\cite{watauchi},
YBa$_{2}$Cu$_{3}$O$_{7-\delta }$ ($\blacklozenge $, $T_{c}\left(
x_{m}\right) =92.9$K, $\gamma _{T_{c}}\left( x_{m}\right) =8$)
\protect\cite{chien123},
YBa$_{2}$(Cu$_{1-y}$Fe$_{y}$)$_{3}$O$_{7-\delta }$ ($\blacksquare
$, $T_{c}\left( x_{m}\right) =92.5$K, $\gamma _{T_{c}}\left(
x_{m}\right) =9$)\protect\cite{chienfe},
Y$_{1-y}$Pr$_{y}$Ba$_{2}$Cu$_{3}$O$_{7-\delta }$
($\blacktriangledown $, $T_{c}\left( x_{m}\right) =91$K, $\gamma
_{T_{c}}\left( x_{m}\right) =9.3$)\protect\cite{chienpr},
BiSr$_{2}$Ca$_{1-y}$Pr$_{y}$Cu$_{2}$O$_{8}$ ($\blacktriangleleft
$, $T_{c}\left( x_{m}\right) =85.4$K, $\gamma _{T=0}\left(
x_{m}\right) =94.3$) \protect\cite{sun} and YBa$_{2}$(Cu$_{1-y}$
Zn$_{y}$)$_{3}$O$_{7-\delta }$ ($\blacktriangleright $,
$T_{c}\left( x_{m}\right) =92.5$K, $\gamma _{T=0}\left(
x_{m}\right) =9$) \protect\cite{panagopzn}. The solid and dashed
curves are Eq.(\ref{eq3}), marking the flow from the maximum
$T_{c}$ to QSI and QSN criticality, respectively.} \label{fig2}
\end{figure}

Provided that this empirical correlation is not merely an artefact
of La$_{2-x}$Sr$_{x}$CuO$_{4}$, it gives a universal perspective
on the interplay of anisotropy and superconductivity, among the
families of cuprates, characterized by $T_{c}\left( x_{m}\right) $
and $\gamma _{T}\left( x_{m}\right) $. For this reason it is
essential to explore its generic validity. In practice, however,
there are only a few additional compounds, including
HgBa$_{2}$CuO$_{4+\delta }$\cite{hoferhg} and
Bi$_{2}$Sr$_{2}$CuO$_{6+\delta }$, for which the dopant
concentration can be varied continuously throughout the entire
doping range. It is well established, however, that the
substitution of magnetic and nonmagnetic impurities, depress
$T_{c}$ of cuprate superconductors very
effectively\cite{xiao,tarascon}. To compare the doping and
substitution driven variations of the anisotropy, we depicted in
Fig. \ref{fig2} the plot $T_{c}/T_{c}\left( x_{m}\right) $ versus
$\gamma _{T}\left( x_{m}\right) /$ $\gamma _{T}$ for a variety of
cuprate families. The collapse of the data on the parabola, which
is the empirical relation (\ref{eq3}), reveals that this scaling
form is well confirmed. Thus, given a family of cuprate
superconductors, characterized by $T_{c}\left( x_{m}\right) $ and
$\gamma _{T}\left( x_{m}\right) $, it gives a generic perspective
on the interplay between anisotropy and superconductivity.

\begin{figure}[tbp]
\centering
\includegraphics[totalheight=6cm]{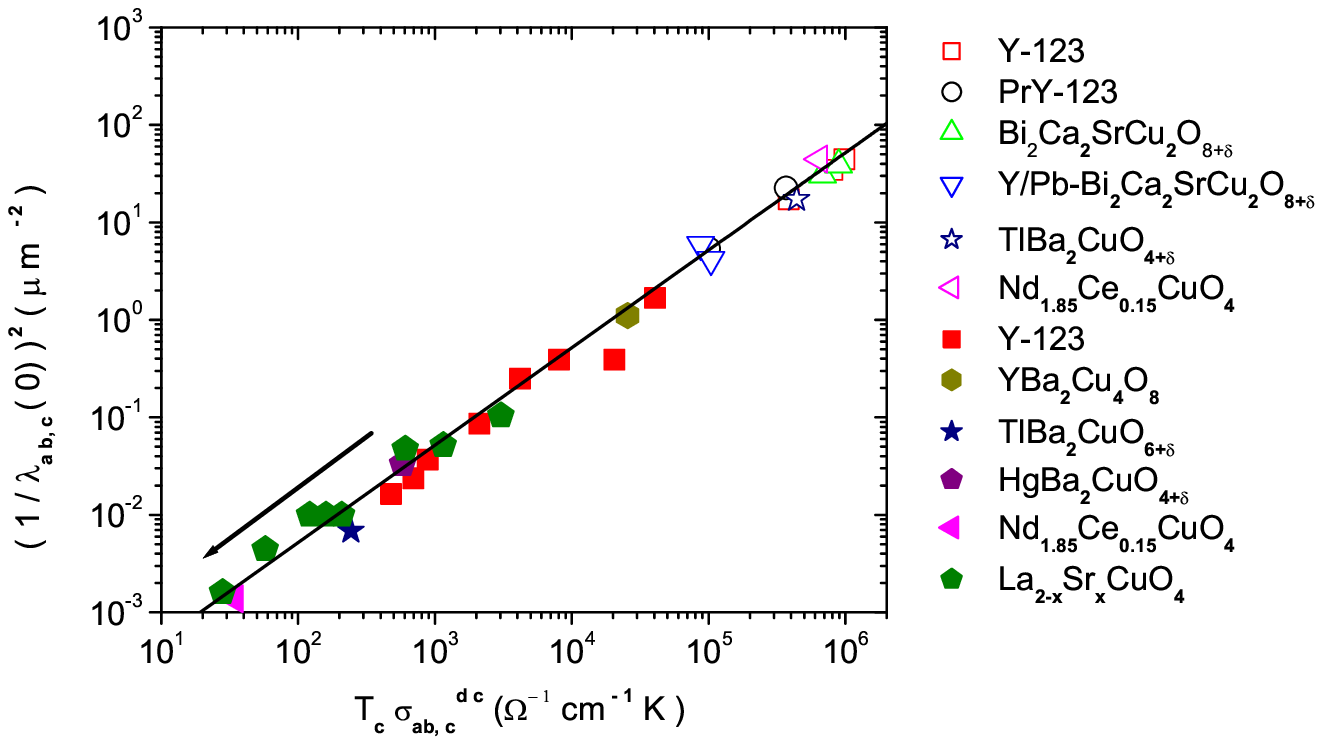}
\caption{$1/\lambda _{ab}^{2}\left( 0\right) $ \textit{vs.}
$T_{c}\sigma _{ab}^{dc}$ (open symbols) and $1/\lambda
_{c}^{2}\left( 0\right) $ \textit{vs.} $T_{c}\sigma _{c}^{dc}$
(full symbols) for various cuprates as collected by Homes \emph{et
al.}\protect\cite{homesuni}. The straight line is $1/\lambda
_{ab,c}^{2}\left( 0\right) =5.2$ $10^{-5}$ $T_{c}\sigma
_{ab,c}^{dc}$. The experimental data is taken from
\protect\cite{10,11,12,19,20,21} for YBa$_{2}$Cu$_{3}$O$_{7-\delta
}$ (Y-123), \protect\cite{12} for Pr-YBa$_{2}$Cu$_{3}$O$_{7-\delta
}$(PrY-123), \protect\cite{19,13} for YBa$_{2}$Cu$_{4}$O$_{8}$,
\protect\cite{19,13} for Tl$_{2}$Ba$_{2}$CuO$_{6+\delta }$,
\protect\cite{12,22} for Bi$_{2}$Ca$_{2}$SrCu$_{2}$O$_{8+\delta }$
and Y/Pb- Bi$_{2}$Ca$_{2}$SrCu$_{2}$O$_{8+\delta }$,
\protect\cite{5,6} for Nd$_{1.85}$Ce$_{0.15}$CuO$_{4}$,
\protect\cite {homesuni,14} for La$_{2-x}$Sr$_{x}$CuO$_{4}$(214),
and \protect\cite{homesuni} for HgBa$_{2}$CuO$_{4+\delta }$. The
arrow indicates the flow to the 2D-QSI critical point.}
\label{fig3}
\end{figure}

Furthermore there is the impressive empirical correlation between
zero temperature penetration depths, transition temperature and
normal state conductivities (extrapolated to $T_{c}$), discovered
by Homes \emph{et al.} \cite{homesuni}. In Fig. \ref{fig3} we
displayed this scaling behavior in terms of $1/\lambda
_{ab}^{2}\left( 0\right) $ \textit{vs}. $T_{c}\sigma _{ab}^{dc}$
and $1/\lambda _{c}^{2}\left( 0\right) $ \textit{vs.} $T_{c}\sigma
_{c}^{dc}$. Here $\sigma _{i}^{dc}$ is the real part of the
frequency dependent normal state conductivity $\sigma
_{i}^{dc}\left( \omega \right) $ in direction $i$ extrapolated to
zero frequency.

Noting that the linear relationship extends over six decades it
provides unprecedented evidence for universal behavior. Indeed,
the data encompasses the slightly underdoped and overdoped regime
of a variety of cuprates. An important implication is that the
changes $\Delta T_{c}$, $\Delta \left( 1/\lambda
_{ab,c}^{2}\right) $ and$\Delta \sigma _{ab,c}^{dc}$, e.g. induced
by isotope exchange, are not independent but related by
\begin{equation}
\frac{\Delta T_{c}}{T_{c}}+2\frac{\Delta \lambda _{ab,c}\left(
0\right) }{\lambda _{ab,c}\left( 0\right) }=-\frac{\Delta \sigma
_{ab,c}^{dc}}{\sigma _{ab,c}^{dc}}.  \label{eq4}
\end{equation}
Such scaling behavior differs drastically from the isotope effects
in the so called conventional superconductors. In these materials
mean-field treatments including the BCS theory apply and for
elemental superconductors $T_{c}$ scales roughly as $M^{-1/2}$,
where $M$ is the mass of the ions. Historically, the resulting
isotope effect on $T_{c}$ identified the phonons as the bosons
mediating superconductivity. Furthermore, the isotope effect on
the penetration depth, entering via the electron-phonon
interaction mediated renormalization of the fermi-velocity,
appears to be negligibly small.

Given the critical line $T_{c}\left( x\right) $ with the 2D-QSI
and 3D-QSN endpoints (see Fig. \ref{fig1}) such universal
relations are not unexpected but a consequence of
fluctuations\cite {book,parks,hohenberg,peliasetto}. As an example
we consider the universal scaling relation at the 2D-QSI
transition\cite{polen,book,klosters,parks,kim},
\begin{equation}
T_{c}=\frac{\Phi _{0}^{2}R_{2}}{16\pi
^{3}k_{B}}\frac{d_{s}}{\lambda _{ab}^{2}\left( 0\right) },
\label{eq5}
\end{equation}
which holds independently of the nature of the putative quantum
critical point. $\lambda _{ab}\left( 0\right) $ is the zero
temperature in-plane penetration depth, $d_{s}$ the thickness of
the sheets and $R_{2}$ is a universal number. Since $T_{c}\propto
d_{s}/\lambda _{ab}^{2}\left( 0\right) \propto n_{s}^{\Box }$,
where $n_{s}^{\Box }$ is the aerial superfluid density, is a
characteristic 2D property, it also applies to the onset of
superfluidity in $^{4}$He films adsorbed on disordered substrates,
where it is well confirmed \cite{crowell}. A great deal of
experimental work has also been done in cuprates on the so called
Uemura plot\cite{uemura}, revealing an empirical correlation
between $T_{c}$ and $d_{s}/\lambda _{ab}^{2}\left( 0\right) $.

\bigskip

This work aims to review the evidence that the empirical scaling
relations between $\lambda _{ab,c}\left( 0\right) $,$T_{c}$,
$d_{s}$, $\sigma _{ab,c}^{dc}$, etc., as well as the resulting
relations for the isotope and pressure effects, reflect 3D-XY
universality along the phase transition line $T_{c}\left( x\right)
$ and the crossover to the 2D-QSI quantum critical point (see Fig.
\ref{fig1}). Although there is considerable evidence that cuprates
fall into the 3D-XY universality class\cite{book,parks,bled,tsdc},
the inhomogeneity induced finite size effects\cite{bled,tsdc} and
the crossovers to 2D-QSI and 3D-QSN criticality make it difficult
to extract critical exponents unambiguously. For this reason we
concentrate on the universal relations between critical
amplitudes, their zero temperature counterparts and $T_{c}$,
because these quantities can be measured with reasonable accuracy.

The paper is organized as follows. In Sec.II we sketch the
universal relations for anisotropic type II superconductors
falling into the 3D-XY universality class and undergo a crossover
to 2D-QSI criticality. Expressing the critical amplitudes of the
penetration depths in terms of their zero temperature
counterparts, we find remarkable agreement with the plot shown in
Fig. \ref{fig3} and the experimental data for
La$_{2-x}$Sr$_{x}$CuO$_{4}$\cite{panagopoulos,uemura,uemura214,komiya,sato}
and Pr$_{2-x}$Ce$_{x}$CuO$_{4-\delta }$\cite{kimpr}, where the
doping dependence of the quantities of interest have been studied.
In this context it should be recognized that the ratio between the
critical amplitudes of the penetration depths and their zero
temperature counterparts is not expected to be universal. However,
the impressive agreement reveals that the doping and family
dependence of this ratio is rather weak. In any case, the
unprecedented consistency of the experimental data with 3D-XY
universality and the crossover to the 2D-QSI critical point,
together with the evidence for the competition between anisotropy
and superconductivity raise again serious doubts whether 2D
mechanisms and models\cite{anderson} can explain these essential
observations.

As the isotope effects are concerned, we confirm previous scaling
relations between the effect on transition temperature and
penetration depths\cite {tshk,tsiso} and derive new relations
including the isotope effect on the conductivity and Hall
constant. Furthermore, the origin of the oxygen isotope effects is
traced back to a change of the c-axis correlation length. It
implies a change of $d_{s}$, a shift of the underdoped limit
($x_{u}$ in Fig. \ref{fig1}) and a reduction of the concentration
of mobile carriers, accompanied by a change of the Hall constant.
Noting that the change of the lattice constants upon oxygen
isotope exchange is negligibly small\cite{conder,raffa} these
isotope shifts reveal the existence and relevance of the coupling
between the superfluid and volume preserving local lattice
distortions. Furthermore, we review the experimental data for the
combined isotope and finite size effects\cite{tsrkhk}. They open a
door to probe the coupling between local lattice distortions and
superconductivity rather directly in terms of the isotope shift of
$L_{c}$, the spatial extent of the homogeneous superconducting
grains along the c-axis. Noting that $\Delta L_{c}/L_{c}$ is
rather large this change confirms the coupling between
superfluidity and local lattice distortion and this coupling is
likely important in understanding the pairing mechanism. Finally
it is shown that the effect of hydrostatic pressure on the
transition temperature and the in-plane penetration depth of
YBa$_{2}$Cu$_{4}$O$_{8}$\cite{khasanovp124} is consistent with
3D-XY universality as well. Thus the remarkable agreement with
3D-XY scaling and the evidence for the crossover to the 2D-QSI
quantum critical point single out 3D and anisotropic microscopic
models which incorporate local lattice distortions, fall in the
experimentally accessible regime into the 3D-XY universality
class, and incorporate the crossover to 2D-QSI criticality where
superconductivity disappears.
\bigskip

\section{Experimental Evidence for the universal 3D-XY- and 2D-QSI- scaling
relations and their implications}

\begin{figure}[tbp]
\centering
\includegraphics[totalheight=6cm]{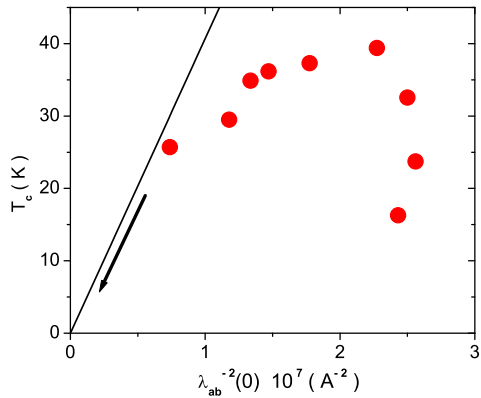}
\caption{$T_{c}$ \textit{vs.} $1/\lambda _{ab}^{2}\left( 0\right)
$ for La$_{2-x}$Sr$_{x}$CuO$_{4}$. Data taken from Uemura \emph{et
al.} \protect\cite {uemura,uemura214} and Panagopoulos \emph{et
al.} \protect\cite{panagopoulos}. The straight line is Eq.
(\ref{eq5}) with $R_{2}d_{s}=6.5$\AA\ and the arrow indicates the
flow to 2D-QSI transition criticality.} \label{fig4}
\end{figure}

It is well-established that strongly type-II materials should
exhibit a continuous normal to superconductor phase transition,
and that sufficiently close to $T_{c}$, the charge of the order
parameter field becomes relevant \cite{mo}. However, in cuprate
superconductors within the fluctuation dominated regime, the
region close to $T_{c}$ where the system crosses over to the
regime of charged fluctuations turns out to be too narrow to
access. For instance, optimally doped
YBa$_{2}$Cu$_{3}$O$_{7-\delta }$, while possessing an extended
regime of critical fluctuations, is too strongly type-II to
observe charged critical fluctuations\cite{book,parks}. Indeed,
the effective dimensionless charge $\widetilde{e}=\xi /\lambda
=1/\kappa $ is in strongly type II superconductors ($\kappa >>1$)
small. The crossover upon approaching $T_{c}$ is thus initially to
the critical regime of a weakly charged superfluid where the
fluctuations of the order parameter are essentially those of an
uncharged superfluid or XY-model \cite{ffh}. Furthermore, there is
the finite size effect due to inhomogeneities which makes the
asymptotic critical regime unattainable\cite{bled,tsdc}. Thus, as
long as the charge of the pairs is negligibly small the cuprates
are expected to fall along the phase transition line $T_{c}\left(
x\right) $ into the 3D-XY universality class. A universality class
is not only characterized by its critical exponents but also by
various critical-point amplitude combinations\cite{hohenberg}. In
particular 3D-XY universality extended to anisotropic
systems\cite{book,parks} then implies that the transition
temperature $T_{c}$ and the critical amplitudes of the penetration
depths $\lambda _{ab0}$ and transverse correlation lengths $\xi
_{ab0}^{tr}$ are not independent but related by\cite{book,parks}
\begin{equation}
k_{B}T_{c}=\frac{\Phi _{0}^{2}}{16\pi ^{3}}\frac{\xi
_{ab0}^{tr}}{\lambda _{ab0}^{2}}=\frac{\Phi _{0}^{2}}{16\pi
^{3}}\frac{\xi _{c0}^{tr}}{\lambda _{c0}^{2}},  \label{eq6}
\end{equation}
where $\lambda _{i}^{2}\left( T\right) =\lambda _{i0}^{2}t^{-\nu
}$ and $\xi _{i}^{t}\left( T\right) =\xi _{i0}^{t}t^{-\nu }$ with
$t=1-T/T_{c}$ and $\nu \simeq 2/3$. For our purpose it is
convenient to replace the transverse correlation length by the
corresponding correlation lengths above $T_{c}$ in terms
of\cite{book,parks,peliasetto}
\begin{equation}
\frac{\xi _{ab0}^{tr}}{\xi _{c0}}=f\approx 0.453,\text{ }\frac{\xi
_{c0}^{tr}}{\xi _{ab0}}=\gamma f,  \label{eq7}
\end{equation}
where the anisotropy is given by
\begin{equation}
\gamma ^{2}=\left( \frac{\lambda _{c0}}{\lambda _{ab0}}\right)
^{2}=\frac{\xi _{c0}^{tr}}{\xi _{ab0}^{tr}}=\left( \frac{\xi
_{ab0}}{\xi _{co}}\right) ^{2}.  \label{eq8}
\end{equation}
Combining Eqs.(\ref{eq6}) and (\ref{eq7}) we obtain the universal
relation
\begin{equation}
T_{c}\lambda _{ab0}^{2}=\frac{\Phi _{0}^{2}f}{16\pi ^{3}k_{B}}\xi
_{c0}. \label{eq9}
\end{equation}
It holds, as long as cuprates fall into the 3D-XY universality
class, irrespective of the doping dependence of $T_{c}$, $\lambda
_{ab0}^{2}$ and $\xi _{c0}$. For this reason it provides a sound
basis for universal plots. However, there is the serious drawback
that reliable experimental estimates for $T_{c}$ and the critical
amplitudes $\lambda _{ab0}$ and $\xi _{c0}$ measured on the same
sample are not yet available. Nevertheless, some progress can be
made by noting that by approaching the 2D-QSI transition the
universal scaling form (\ref{eq9}) should match with Eq.
(\ref{eq5}). This requires
\begin{equation}
\lambda _{ab0}=\Lambda _{ab}\lambda _{ab}\left( 0\right) ,\text{
}f\xi _{c0}/\Lambda _{ab}^{2}\rightarrow R_{2}d_{s},  \label{eq10}
\end{equation}
so that away from 2D-QSI criticality
\begin{equation}
T_{c}\lambda _{ab}^{2}\left( 0\right) \simeq \frac{\Phi
_{0}^{2}f}{16\pi ^{3}k_{B}}\frac{\xi _{c0}}{\Lambda _{ab}^{2}},
\label{eq11}
\end{equation}
holds. In Fig. \ref{fig4} we displayed $T_{c}$ \textit{vs.}
$1/\lambda _{ab}^{2}\left( 0\right) $ for
La$_{2-x}$Sr$_{x}$CuO$_{4}$ taken from Uemura \emph{et al.}
\cite{uemura,uemura214} and Panagopoulos \emph{et al.} \cite
{panagopoulos}. The straight line is Eq. (\ref{eq5}) with
$R_{2}d_{s}=6.5$\AA\ and the arrow indicates the flow to 2D-QSI
transition criticality. Thus, when both $T_{c}$ and $1/\lambda
_{ab}^{2}\left( 0\right) $ increase, $T_{c}$ values below
$T_{c}=\left( \Phi _{0}^{2}R_{2}/6\pi ^{3}k_{B}\right)
d_{s}/\lambda _{ab}^{2}\left( 0\right) $ (Eq. (\ref{eq5})) require
$\xi _{c0}$ to fall off from its limiting value $\xi
_{c0}=d_{s}R_{2}\Lambda _{ab}^{2}/f$.

\begin{figure}[tbp]
\centering
\includegraphics[totalheight=6cm]{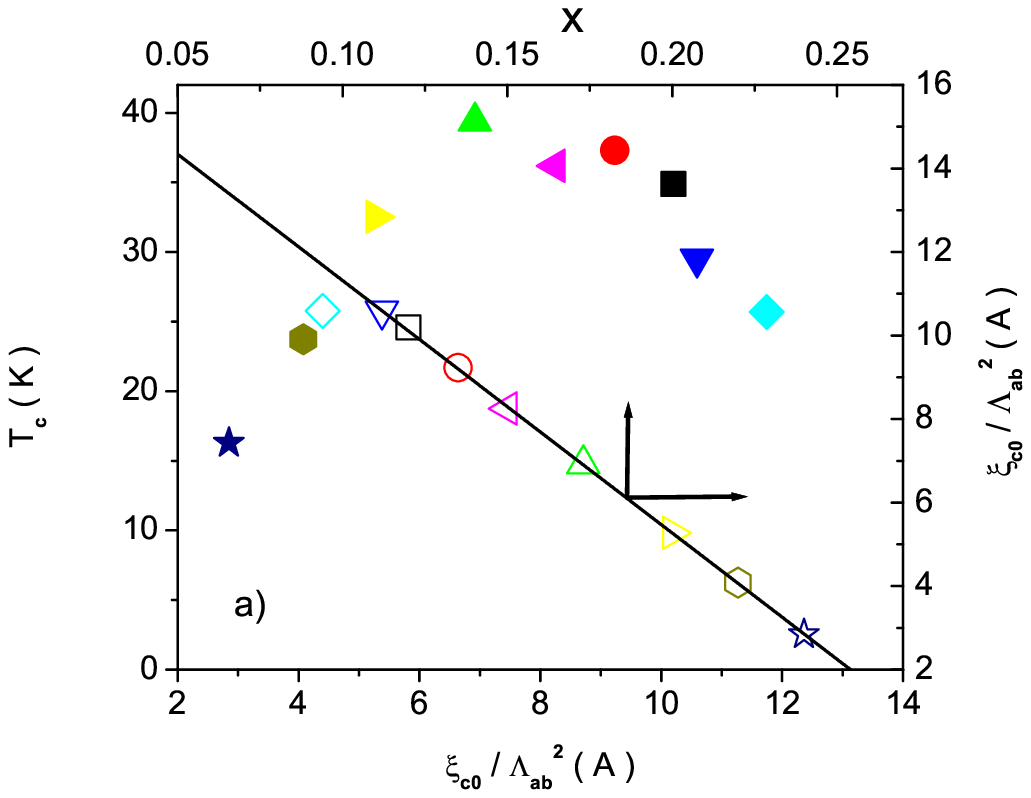}
\includegraphics[totalheight=6cm]{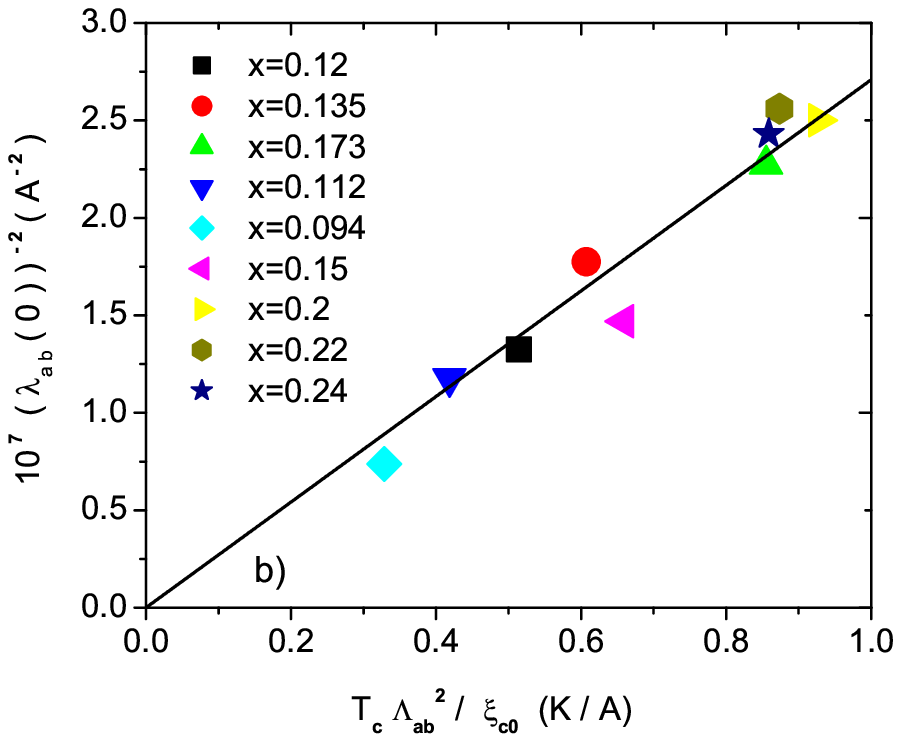}
\caption{a) $T_{c}\lambda _{ab}^{2}\left( 0\right) \propto \xi
_{c0}/\Lambda _{ab}^{2}\ $\textit{vs.} $T_{c}$ and $T_{c}\lambda
_{ab}^{2}\left( 0\right) \propto \xi _{c0}/\Lambda _{ab}^{2}$
\textit{vs. }$x$ for La$_{2-x}$Sr$_{x}$CuO$_{4}$. Data taken from
Pangopoulos \emph{et al.} \protect\cite{panagopoulos} and Uemura
\emph{et al.} \protect\cite{uemura214}. The solid line indicates
the behavior in the underdoped and the dashed one in the overdoped
regime, while the dotted line corresponds to Eq. (\ref{eq12}) in
the form $T_{c}\lambda _{ab}^{2}\left( 0\right) =37-156(x-0.047)$.
b) $1/\lambda _{ab}^{2}\left( 0\right) \ $\textit{vs.} $\Lambda
_{ab}^{2}T_{c}/\xi _{c0}$ for the same data with $\xi
_{c0}/\Lambda _{ab}^{2}$ given by Eq. (\ref{eq12}). The straight
line is $1/\lambda _{ab}^{2}\left( 0\right) =2.71$ $T_{c}\Lambda
_{ab}^{2}/\xi _{c0}$. } \label{fig5}
\end{figure}

The doping dependence of $\xi _{c0}/\Lambda _{ab}^{2}$ deduced
from Eq. (\ref {eq11}) and the experimental data for $T_{c}$ and
$\lambda _{ab}^{2}\left( 0\right) $ is displayed in Fig.
\ref{fig5}a in terms of $T_{c}\lambda _{ab}^{2}\left( 0\right)
\propto \xi _{c0}/\Lambda _{ab}^{2}\ $\textit{vs.} $ T_{c}$ and
$T_{c}\lambda _{ab}^{2}\left( 0\right) \propto \xi _{c0}/\Lambda
_{ab}^{2}$ \textit{vs. }$x$ . Approaching the underdoped limit
$\left( x\approx 0.05\right) $, where $T_{c}$ vanishes (Fig.
\ref{fig1}) and the 2D-QSI transition occurs, $\xi _{c0}/\Lambda
_{ab}^{2}$ increases nearly linearly with decreasing $x$ to
approach a fixed value. Indeed, the data is consistent with
\begin{equation}
\xi _{c0}/\Lambda _{ab}^{2}=\left( 16\pi ^{3}k_{B}/\left( \Phi
_{0}^{2}f\right) \right) T_{c}\lambda _{ab}^{2}\left( 0\right)
\approx 14.34-60.47(x-0.05)\text{\AA },  \label{eq12}
\end{equation}
yielding the limiting value $\xi _{c0}\left( x=0.05\right)
/\Lambda _{ab}^{2}\approx 14.34\text{\AA\ and }f\xi _{c0}\left(
x=0.05\right) /\Lambda _{ab}^{2}=R_{2}d_{s}\approx 6.5$\AA\ used
in Fig. \ref{fig4}. An essential result is that $\xi _{c0}$ adopts
in the underdoped limit ($x\simeq 0.05$) where the 2D-QSI
transition occurs and $T_{c}$ vanishes its maximum value $\Lambda
\xi _{c0}/\Lambda _{ab}^{2}\approx 14.34$\AA , which is close to
the c-axis lattice constant $c\simeq 13.29$\AA . Thus, a finite
transition temperature requires a reduction of $\xi _{c0}$, well
described over an unexpectedly large doping range by Eq.
(\ref{eq12}). with the doping dependence of $\gamma $ (Eq.
(\ref{eq2})) Eq. (\ref{eq8}) transforms to
\begin{equation}
\xi _{c0}/\Lambda _{ab}^{2}\approx 14.34-60.47\gamma _{0}/\gamma
\text{\AA }, \label{eq13}
\end{equation}
revealing that the doping dependence of \ the c-axis correlation
length $\xi _{c0}$ is intimately related to the anisotropy $\gamma
$. Hence, a finite $T_{c}$ requires unavoidably a finite
anisotropy $\gamma $.

\begin{figure}[tbp]
\centering
\includegraphics[totalheight=6cm]{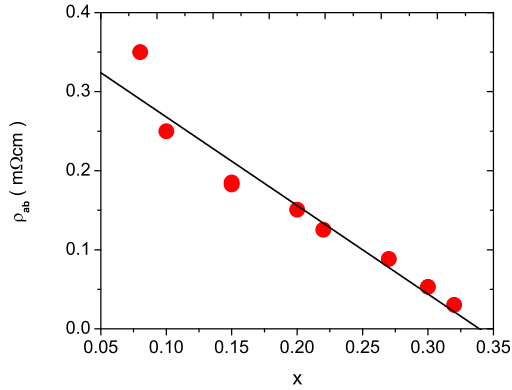}
\caption{$\rho _{ab}$ \textit{vs.} $x$ for
La$_{2-x}$Sr$_{x}$CuO$_{4}$ derived from the data of Komiya
\emph{et al.}\protect\cite{komiya} and Sato \emph{et al.}
\protect\cite{sato}. $\rho _{ab}$ is determined above $T_{c}$ from
$\rho _{ab}\left( T\right) $ in terms of a linear extrapolation
yielding $\rho _{ab}=\rho _{ab}\left( T_{c}^{+}\right) $. The
solid line indicates the consistency with a nearly linear doping
dependence.} \label{fig6}
\end{figure}

The lesson is, that superconductivity in
La$_{2-x}$Sr$_{x}$CuO$_{4}$ is an anisotropic but 3D phenomenon
which disappears in the 2D limit. From the plot $1/\lambda
_{ab}^{2}\left( 0\right) \ $\textit{vs.} $\Lambda
_{ab}^{2}T_{c}/\xi _{c0}$ displayed in Fig. \ref{fig5}b, where
according to Eq.(\ref{eq11}) universal behavior is expected to
occur, the data is seen to fall rather well on a straight line.
This is significant, as moderately underdoped, optimally and
overdoped La$_{2-x}$Sr$_{x}$CuO$_{4}$ falling according to Fig.
\ref{fig4} well off the 2D-QSI behavior $T_{c}\propto 1/\lambda
_{ab}^{2}\left( 0\right) $ now scale nearly onto a single line.
Thus the approximate 3D-XY scaling relation (\ref{eq11}), together
with the empirical doping dependence of the c-axis correlation
length (Eqs. (\ref{eq12}) and (\ref{eq13})) are consistent with
the available experimental data for La$_{2-x}$Sr$_{x}$CuO$_{4}$
and uncovers the relevance of the anisotropy. However, the linear
doping dependence of $\xi _{c}$ is not expected to hold closer to
the overdoped limit $\left( x\approx 0.27\right) $ where a 3D
quantum superconductor to normal state (3D-QSN) transition is
expected to occur\cite {book,parks}. Clearly, a full test of the
scaling relation (\ref{eq11}) requires independent experimental
data for the critical amplitude $\xi _{ab0}$ of the in-plane
correlation length. To identify the observable probing this
correlation length we note that close to 2D-QSI criticality the
sheet conductivity and conductivity are related by $\sigma
_{sheet}=d_{s}\sigma _{ab}^{dc}$ so that the universal relation
(\ref{eq5}) can be rewritten in the form
\begin{equation}
\frac{1}{\lambda _{ab}^{2}\left( 0\right) T_{c}\sigma
_{ab}^{dc}}=\frac{16\pi ^{3}k_{B}}{\Phi _{0}^{2}R_{2}\sigma
_{sheet}},\text{ \ }\sigma _{sheet}=\frac{h}{4e^{2}}\sigma
_{0}\simeq \sigma _{0}\text{ }1.55\text{ }10^{-4}\Omega ^{-1},
\label{eq14}
\end{equation}
where $h/4e^{2}=6.45$k$\Omega $ is the quantum of resistance and
$\sigma _{0} $ is a dimensionless constant of order
unity\cite{herbutsig}. Away from 2D-QSI criticality $\xi _{c0}$
can be expressed as
\begin{equation}
\text{ }1/\xi _{c0}\simeq \sigma _{ab}/s_{ab}=1/\left( \rho
_{ab}s_{ab}\right) ,  \label{eq15}
\end{equation}
with $\rho _{ab}$ determined from $\rho _{ab}\left( T\right) $
above $T_{c}$ by linear extrapolation yielding $\rho _{ab}=\rho
_{ab}\left( T_{c}^{+}\right) $. Up to the non-universal factor
$s_{ab}\left( \Omega ^{-1}\right) $ this resistivity is expected
to reflect the doping dependence of the critical amplitude $\xi
_{c0}$. To demonstrate this behavior we displayed in Fig.
\ref{fig6} $\rho _{ab}$ \textit{vs.} $x$ for
La$_{2-x}$Sr$_{x}$CuO$_{4}$ derived from the data of Komiya
\emph{et al.}\cite{komiya} and Sato \emph{et al.} \cite{sato}.
$\rho _{ab}$ is derived from $\rho _{ab}\left( T\right) $ by
linear extrapolation yielding $\rho _{ab}=\rho _{ab}\left(
T_{c}^{+}\right) $. The solid line points to a nearly linear
doping dependence, consistent with the behavior shown in Fig.
\ref{fig5}a, derived from $T_{c}\lambda _{ab}^{2}\left( 0\right)
\propto \xi _{c0}/\Lambda _{ab}^{2}$ \textit{vs. }$x$.
Consequently, according to Eq. (\ref{eq11}) data for $T_{c}$,
$\lambda _{ab}^{2}\left( 0\right) $ and $\rho _{ab}$, measured on
the same sample at various dopant concentrations, should then
scale as
\begin{equation}
\frac{1}{\lambda _{ab}^{2}\left( 0\right) T_{c}\sigma
_{ab}}=\frac{\rho _{ab}}{\lambda _{ab}^{2}\left( 0\right)
T_{c}}\simeq \frac{16\pi ^{3}k_{B}\Lambda _{ab}^{2}}{\Phi
_{0}^{2}fs_{ab}}.  \label{eq16}
\end{equation}
Thus, in the plot $1/\lambda _{ab}^{2}\left( 0\right) $
\textit{vs. }$T_{c}\sigma _{ab}$ or $T_{c}/\rho _{ab}$ the data
should tend to fall on a straight line, while deviations are
attributable to the non-universal nature of the ratio $\Lambda
_{ab}^{2}/s_{ab}$. Indeed $\Lambda _{ab}^{2}/s_{ab}$ does not
necessarily adopt a unique value.

\begin{figure}[tbp]
\centering
\includegraphics[totalheight=6cm]{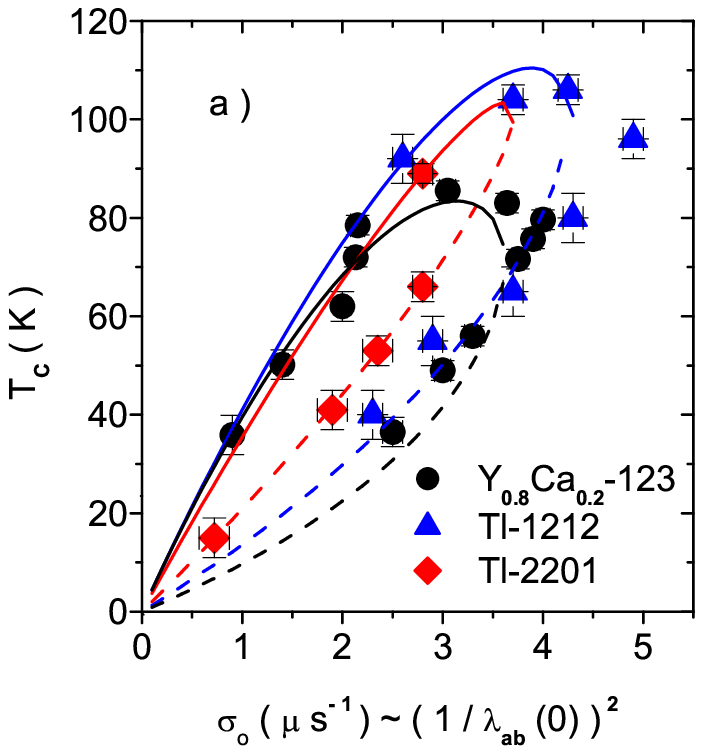}
\includegraphics[totalheight=6cm]{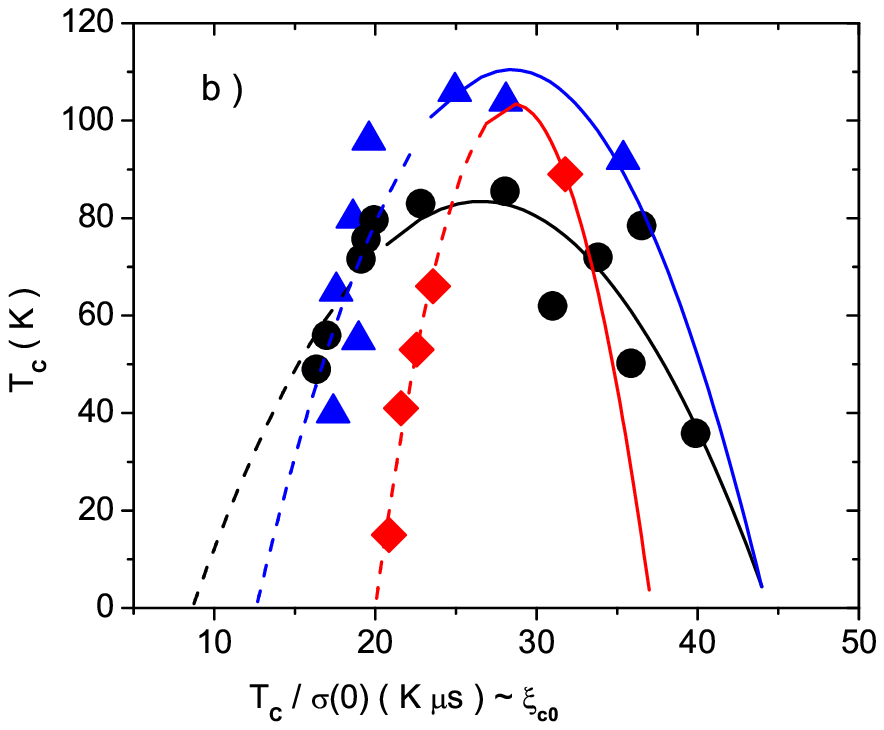}
\caption{a)\ $T_{c}$ \emph{vs.} $\sigma \left( 0\right) \propto
1/\lambda _{ab}^{2}\left( 0\right) $ for Y$_{0.8}$Ca$_{0.2}$-123,
Tl-1212 and Tl-2201 taken from Bernhard \emph{et
al}.\protect\cite{bernhard} and Niedermayer \emph{et
al.}\protect\cite{niedermayer}. The solid curves indicate the flow
from optimum doping to 2D-QSI criticality and the dashed ones the
crossover to the 3D-QSN transition in the overdoped limit. b)\
$T_{c}$ \emph{vs.} $T_{c}/\sigma \left( 0\right) \propto
T_{c}\lambda _{ab}^{2}\left( 0\right) \propto $ $\xi _{c0}$ for
the same data. The solid lines indicate the underdoped and the
dashed ones the overdoped regime.} \label{fig7}
\end{figure}

To demonstrate that this behavior is not an artefact of
La$_{2-x}$Sr$_{x}$CuO$_{4}$ we displayed in Fig. \ref{fig7}a the
$\mu $SR data $T_{c}$ \emph{vs.} $\sigma \left( 0\right) \propto
1/\lambda _{ab}^{2}\left( 0\right) $ for
Y$_{0.8}$Ca$_{0.2}$Ba$_{2}$(Cu$_{1-y}$Zn$_{y}$)O$_{7-\delta
}$(Y$_{0.8}$Ca$_{0.2}$-123),
Tl$_{0.5-y}$Pb$_{0.5+y}$Sr$_{2}$Ca$_{1-x}$Y$_{x}$Cu$_{2}$O$_{7}
$(Tl-1212) and TlBa$_{2}$CuO$_{6+\delta }$(Tl-2201) taken from
Bernhard \emph{et al}.\cite{bernhard} and Niedermayer \emph{et
al.}\cite{niedermayer}. Noting that in these materials the
relationship between $T_{c}$, $1/\lambda _{ab}^{2}\left( 0\right)
$ and the dopant concentration is much less obvious, we displayed
in Fig. \ref{fig7}b the behavior of the critical amplitude of the
c-axis correlation length in terms of $T_{c}$ \emph{vs.}
$T_{c}/\sigma \left( 0\right) \propto T_{c}\lambda _{ab}^{2}\left(
0\right) \propto $ $\xi _{c0}$. Noting that in analogy to
La$_{2-x}$Sr$_{x}$CuO$_{4}$ (see Fig. \ref{fig4}) $T_{c}$
\emph{vs.} $\sigma \left( 0\right) \propto 1/\lambda
_{ab}^{2}\left( 0\right) $ resembles the outline of a fly's wing,
it becomes evident that the behavior of $T_{c}$ \emph{vs.} $\xi
_{c0}$, resembling the doping dependence of $T_{c}$ in
La$_{2-x}$Sr$_{x}$CuO$_{4}$ (see Fig. \ref{fig5}a), appears to be
generic. At the 2D-QSI transition $\xi _{c0}$ adopts a finite
value $\xi _{c0}=d_{s}R_{2}\Lambda _{ab}^{2}/f$, it decreases with
increasing transition temperature and decreases further as the
maximum transition temperature is passed.

\begin{figure}[tbp]
\centering
\includegraphics[totalheight=6cm]{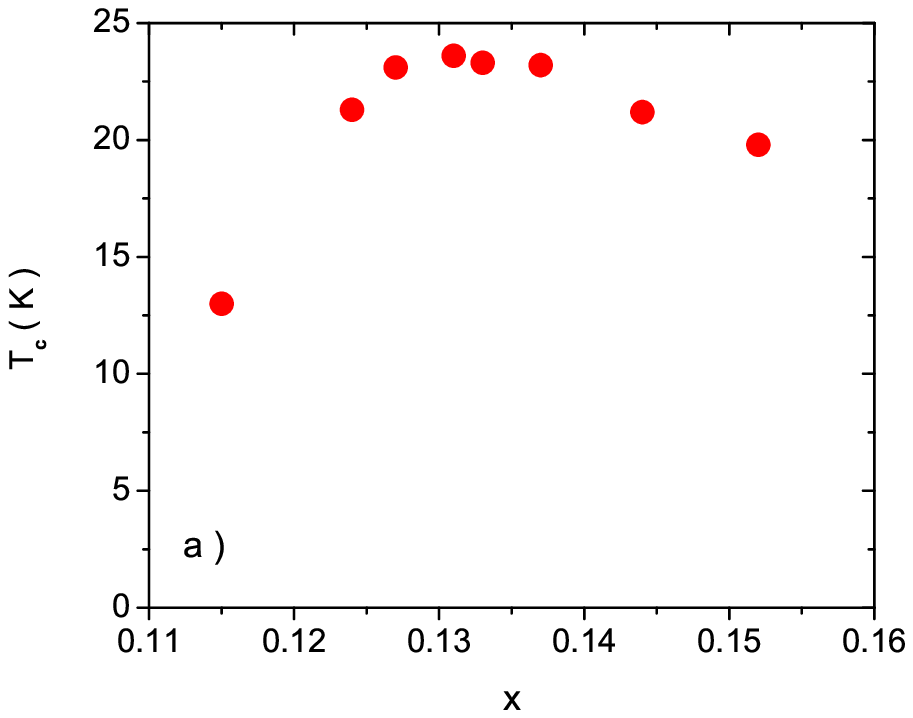}
\includegraphics[totalheight=6cm]{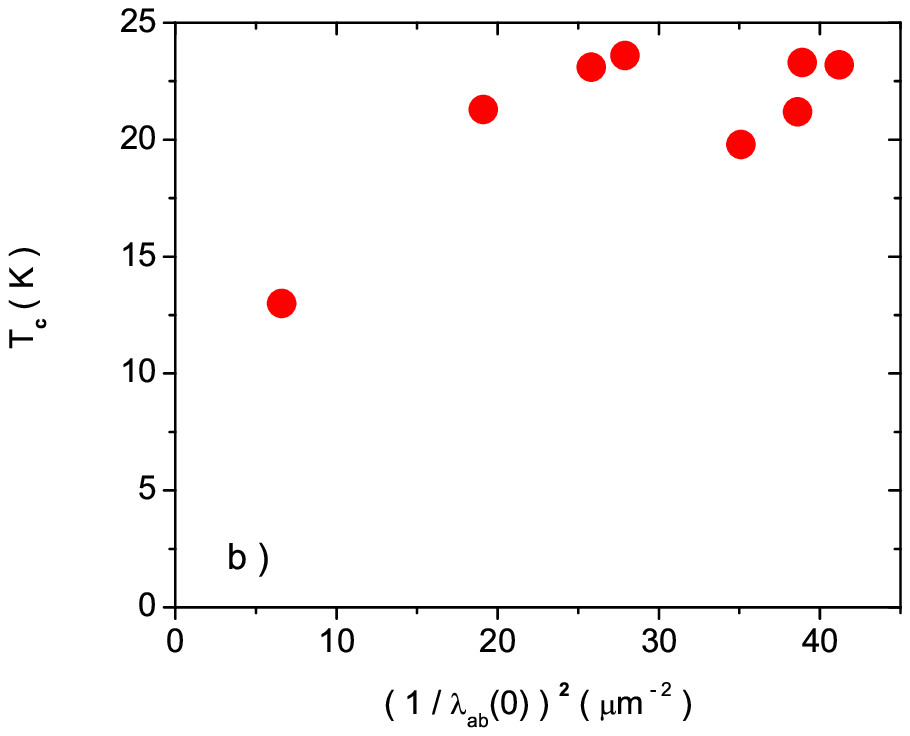}
\caption{a) $T_{c}$ \emph{vs.} $x$ for
Pr$_{2-x}$Ce$_{x}$CuO$_{4-\delta }$; b) $T_{c}$ \emph{vs.}
$1/\lambda _{ab}^{2}\left( 0\right) $. Data taken from Kim
\emph{et al.}\protect\cite{kimpr}.} \label{fig8}
\end{figure}

Further evidence for this behavior stems from the in-plane
penetration and resistivity data of Kim \emph{et al.}\cite{kim}for
the n-type cuprate Pr$_{2-x}$Ce$_{x}$CuO$_{4-\delta }$, which
extends over the range $0.124\leq x\leq 0.144$. From Fig.
\ref{fig8}a it is seen that in analogy to p-doping there is an
insulator to superconductor quantum transition at some $x_{u}$.
For $x>x_{u}$ $T_{c}$ increases monotonically, adopts its maximum
value at optimum p-doping and decreases in the overdoped regime.
Given this analogy between the doping dependence of $T_{c}$ of p-
and n-type cuprates one expects that $T_{c}$ \emph{vs.} $1/\lambda
_{ab}^{2}\left( 0\right) $ uncovers essentially the behavior of
the p-type cuprates shown in Figs. \ref {fig4} and \ref{fig7}a. A
glance to Fig. \ref{fig8}b shows that this indeed the case.
Accordingly, Pr$_{2-x}$Ce$_{x}$CuO$_{4-\delta }$ is expected to
undergo a 2D-QSI transition at some $x_{u}$.

\begin{figure}[tbp]
\centering
\includegraphics[totalheight=6cm]{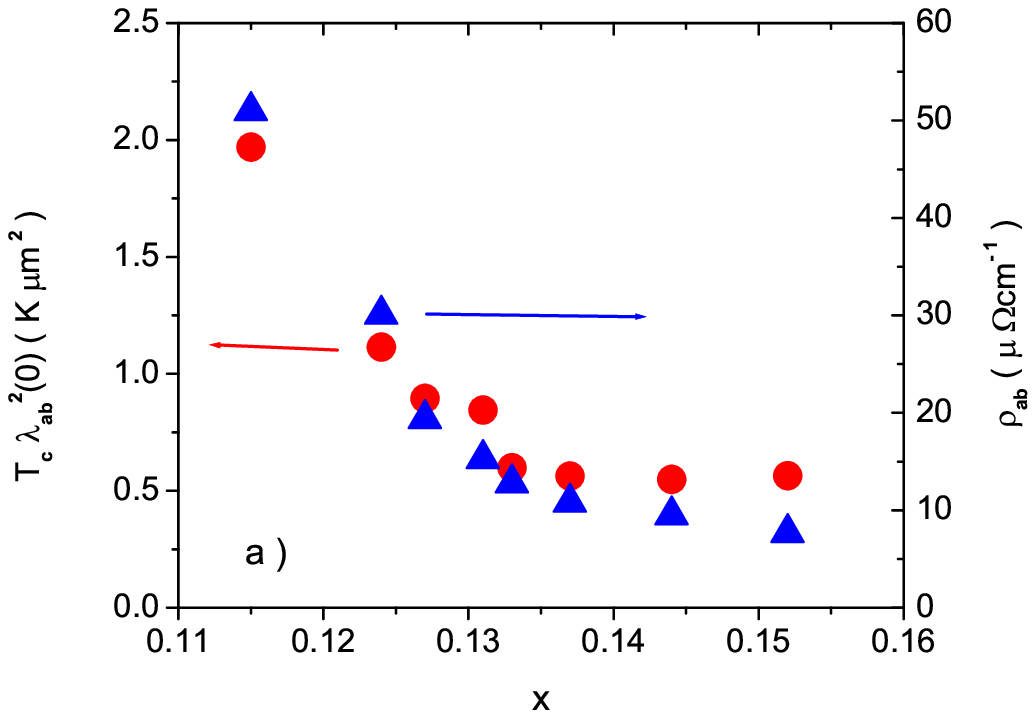}
\includegraphics[totalheight=6cm]{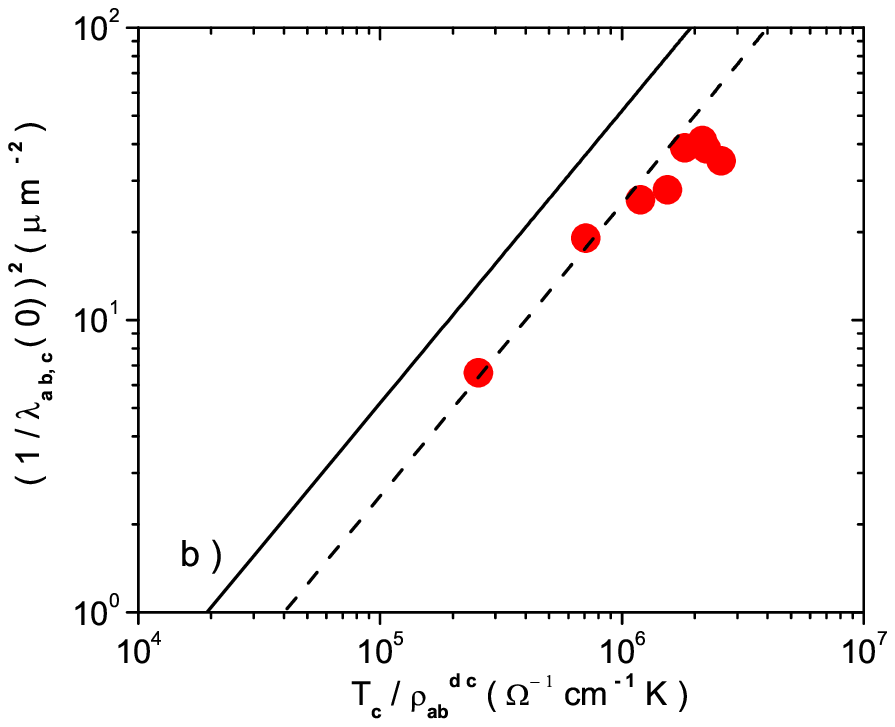}
\caption{a) $\xi _{c0}\propto T_{c}\lambda _{ab}^{2}\left(
0\right) $ $\mathit{vs}\emph{.}$ $x$ $\ $and\ $\ \ \ \rho _{ab}\
\mathit{vs}\emph{.}$ $x\ $\ for Pr$_{2-x}$Ce$_{x}$CuO$_{4-\delta
}$ derived from the data of Kim \emph{et al.}\protect\cite{kim};
b) $1/\lambda _{ab}^{2}\left( 0\right) $ $\mathit{vs}\emph{.}$
$T_{c}/\rho _{ab}$ for the same data. The solid lines line is $
1/\lambda _{ab}^{2}\left( 0\right) =5.2$ $10^{-5}$ $T_{c}/\rho
_{ab}$ and the dashed one $1/\lambda _{ab}^{2}\left( 0\right)
=2.5$ $10^{-5}$ $T_{c}/\rho _{ab}$.} \label{fig9}
\end{figure}

Indeed the doping dependence of $\xi _{c0}\propto T_{c}\lambda
_{ab}^{2}\left( 0\right) $ displayed in Fig. \ref{fig9}a is rather
analogous to that in La$_{2-x}$Sr$_{x}$CuO$_{4}$ shown in Fig.
\ref{fig5}a and $\xi _{c0}$ is seen to scale as $\rho _{ab}$, in
agreement with Eq. (\ref{eq16}). In this view it is not surprising
that in the plot $1/\lambda _{ab}^{2}\left( 0\right) $
$\mathit{vs}\emph{.}$ $T_{c}/\rho _{ab}$ shown in Fig.\ref{fig9}b
the data fall nearly on a straight line. The solid and dashed
lines will be discussed later.

In any case, since Eq. (\ref{eq9}) is universal, it holds for all
cuprates falling in the accessible critical regime into the 3D-XY
universality class, irrespective of the doping dependence of
$T_{c}$, $\lambda _{ab0}^{2}$, $\gamma $ and $\xi _{c0}$. Since
charged criticality is accessible in the heavily underdoped regime
only\cite{tsrkhkcharge}, Eq. (\ref{eq9}) rewritten in the form
\begin{equation}
\frac{\xi _{c0}}{\lambda _{ab0}^{2}T_{c}}=\frac{\xi _{ab0}\gamma
_{T_{c}}}{\lambda _{c0}^{2}T_{c}}=\frac{16\pi ^{3}k_{B}}{\Phi
_{0}^{2}f}, \label{eq17}
\end{equation}
provides a sound basis for universal plots. However, as
aforementioned there is the serious drawback that reliable
experimental estimates for the critical amplitudes and the
anisotropy at $T_{c}$ measured on the same sample and for a
variety of cuprates are not yet available. Nevertheless, as in the
case of La$_{2-x}$Sr$_{x}$CuO$_{4}$ outlined above, progress can
be made by invoking the approximate scaling form (\ref{eq11}), by
expressing the critical amplitude of the penetration depths by
there zero temperature counterparts (Eq.(\ref{eq10})). The
universal relation (\ref{eq17}) transforms then to
\begin{equation}
\frac{\xi _{c0}}{\lambda _{ab}^{2}\left( 0\right) T_{c}}\simeq
\frac{\gamma _{T_{c}}\xi _{ab0}}{\lambda _{c}^{2}\left( 0\right)
T_{c}}\frac{\Lambda _{ab}^{2}}{\Lambda _{c}^{2}}\simeq \frac{16\pi
^{3}k_{B}\Lambda _{ab}^{2}}{\Phi _{0}^{2}f},\text{ }\frac{\Lambda
_{c}}{\Lambda _{ab}}=\frac{\gamma _{T=0}}{\gamma _{T_{c}}},
\label{eq18}
\end{equation}
which will be used to explore the relationship between the isotope
effects on the transition temperature and the zero temperature
penetration depths. As aforementioned a full test of this scaling
form requires an independent experimental determination of the
critical amplitudes of the correlation lengths $\xi _{ab0}$ and
$\xi _{c0}$. We have seen that this is achieved in terms of the
conductivity. Introducing $\sigma _{i}^{dc}$, the real part of the
frequency dependent conductivity $\sigma _{i}^{dc}\left( \omega
\right) $ in direction $i$ extrapolated to zero frequency at
$T\gtrsim T_{c}$\cite {homesuni}, one obtains in analogy to Eq.
(\ref{eq11})
\begin{equation}
\text{ }1/\xi _{c0}\simeq \sigma _{ab}^{dc}/s_{ab},\text{
}1/\left( \gamma _{T_{c}}\xi _{ab0}\right) \simeq \sigma
_{c}^{dc}/s_{c},  \label{eq19}
\end{equation}
with $s_{i}$ in units $\Omega ^{-1}$. With that the relation
(\ref{eq18}) transforms to
\begin{equation}
\frac{1}{\lambda _{ab}^{2}\left( 0\right) T_{c}\sigma
_{ab}^{dc}}\simeq \frac{1}{\lambda _{c}^{2}\left( 0\right)
T_{c}\sigma _{c}^{dc}}\frac{\Lambda _{ab}^{2}}{\Lambda
_{c}^{2}}\simeq \frac{16\pi ^{3}k_{B}\Lambda _{ab}^{2}}{\Phi
_{0}^{2}fs_{ab}},\text{ \ }\frac{\Lambda _{c}}{\Lambda
_{ab}}=\frac{\gamma _{T_{c}}}{\gamma _{T=0}},  \label{eq20}
\end{equation}
because $\sigma _{ab}^{dc}/\sigma _{c}^{dc}=\gamma ^{2}$ and with
that $s_{ab}=s_{c}$. To check this relation we note that the
2D-QSI scaling form (\ref{eq14}) yields
\begin{equation}
\frac{1}{\lambda _{ab}^{2}\left( 0\right) T_{c}\sigma
_{ab}^{dc}}\simeq \frac{10.3\text{ }10^{-5}}{R_{2}\sigma _{0}},
\label{eq21}
\end{equation}
with $\lambda _{ab}\left( 0\right) $ in $\mu $m, $T$ in K and
$\sigma _{ab}^{dc}$ in ($\Omega ^{-1}$cm$^{-1}$), the structure of
the $ab$-expression is recovered.

\begin{figure}[tbp]
\centering
\includegraphics[totalheight=6cm]{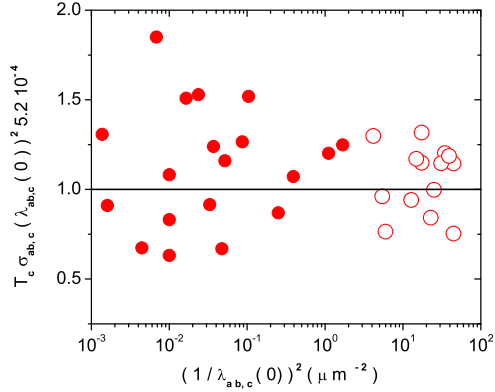}
\caption{$\lambda _{ab,c}^{2}\left( 0\right) T_{c}\sigma
_{ab,c}^{dc}$ $5.210^{-4}$ \textit{vs.} $1/$ $\lambda
_{ab,c}^{2}\left( 0\right) $ for the data shown in Fig.
\ref{fig5}. $\bullet $ : c-axis; $\bigcirc $: $ab$-plane.}
\label{fig10}
\end{figure}

In Fig. \ref{fig3} we displayed $1/\lambda _{ab}^{2}\left(
0\right) $ \textit{vs}. $T_{c}\sigma _{ab}^{dc}$ and $1/\lambda
_{c}^{2}\left( 0\right) $ \textit{vs.} $T_{c}\sigma _{c}^{dc}$ as
collected by Homes \emph{et al.} \cite{homesuni}. Noting that the
linear relationship extends over six decades this plot represents
unprecedented evidence that all these cuprates fall into the 3D-XY
universality class, where the approximate scaling form
(\ref{eq18}) applies. Since the $ab$-plane and $c$-axis data are
well described by the same line, namely $1/\lambda
_{ab,c}^{2}\left( 0\right) \simeq 5.2\ 10^{-5}T_{c}\sigma
_{ab,c}^{dc}$, it follows that $\Lambda _{ab}^{2}/\Lambda
_{c}^{2}$ is close to one and $\Lambda _{ab}^{2}/s_{ab}$ adopts a
nearly unique value. Note that such line is also consistent with
the 2D-QSI scaling form (\ref{eq21}). In particular for
$R_{2}\sigma _{0}\cong 1.98$ these lines coincide. The fact that
all points of $1/\lambda _{ab}^{2}\left( 0\right) $ \textit{vs.}
$T_{c}\sigma _{ab}^{dc}$ (open symbols) nearly fall onto a single
line is significant, as moderately underdoped, optimally and
overdoped materials, which fell well off of the 2D-QSI behavior
$T_{c}\propto 1/\lambda _{ab}^{2}\left( 0\right) $ (see Figs.
\ref{fig4}, \ref{fig7}a, and \ref{fig8}a) now scale nearly onto a
single line, in agreement with Figs. \ref{fig5}b and \ref{fig9}b.
Although the agreement, extending over 6 decades in the scaling
variables is impressive, the plot $\lambda _{ab,c}^{2}\left(
0\right) T_{c}\sigma _{ab,c}^{dc}$ $5.210^{-4}$ \textit{vs.} $1/$
$\lambda _{ab,c}^{2}\left( 0\right) $ displayed in Fig.
\ref{fig10} reveals that there are deviations, ascribable to
experimental uncertainties and the fact that $\Lambda
_{ab}^{2}/\Lambda _{c}^{2}$ and $\Lambda _{ab}^{2}/s_{ab}$ are
non-universal quantities of order 1. The non-universality of
$\Lambda _{ab}^{2}/s_{ab}$ clearly emerges from the data for
Pr$_{2-x}$Ce$_{x}$CuO$_{4-\delta }$ displayed in Fig. \ref{fig9}b,
pointing to $1/\lambda _{ab}^{2}\left( 0\right) =2.5$ $10^{-5}$
$T_{c}/\rho _{ab}$ (dashed line) in contrast to $1/\lambda
_{ab}^{2}\left( 0\right) =5.2$ $10^{-5}$ $T_{c}/\rho _{ab}$ (solid
line).

In any case, the overall impressive evidence for anisotropic 3D-XY
scaling raises again serious doubts that 2D models\cite{anderson}
are potential candidates to explain superconductivity in the
cuprates. Indeed, the doping or $T_{c\text{ }}$dependence of the
c-axis correlation length is an essential ingredient (see Figs.
\ref{fig5}a, Fig. \ref{fig7}b and Fig. \ref {fig9}a). Furthermore,
since both, $1/\lambda _{ab}^{2}\left( 0\right) $ and $1/\lambda
_{c}^{2}\left( 0\right) $ tend to zero by approaching 2D-QSI
criticality, this plot also uncovers the flow to this quantum
critical point, as indicated by the arrow in Fig. \ref{fig3}.

Having established the remarkable reliability of the scaling
relations (\ref {eq18}) and (\ref{eq20}) we are now prepared to
discuss the isotope effects on $T_{c}$ and the zero temperature
in-plane penetration depth. Since Eq. (\ref{eq5}) is universal, it
also implies that the changes $\Delta T_{c}$, $\Delta d_{s}$ and
$\Delta \left( 1/\lambda _{ab}^{2}\left( T=0\right) \right) $,
induced by pressure or isotope exchange are not independent, but
close to 2D-QSI criticality related by\cite{tshk}
\begin{equation}
\frac{\Delta T_{c}}{T_{c}}+2\frac{\Delta \left( \lambda
_{ab}\left( 0\right) \right) }{\lambda _{ab}\left( 0\right)
}=\frac{\Delta d_{s}}{d_{s}}, \label{eq22}
\end{equation}
while the approximate scaling relation (\ref{eq18}), applicable
over an extended doping range (see Fig. \ref{fig3}) yields
\begin{equation}
\frac{\Delta T_{c}}{T_{c}}+2\frac{\Delta \left( \lambda
_{ab}\left( 0\right) \right) }{\lambda _{ab}\left( 0\right)
}=\frac{\Delta \left( \xi _{c0}/\Lambda _{ab}^{2}\right) }{\left(
\xi _{c0}/\Lambda _{ab}^{2}\right) }, \label{eq23}
\end{equation}
where according to relation (\ref{eq18})
\begin{equation}
\frac{\xi _{c0}}{\Lambda _{ab}^{2}}=\frac{16\pi ^{3}k_{B}}{\Phi
_{0}^{2}f}\lambda _{ab}^{2}\left( 0\right) T_{c}.  \label{eq24}
\end{equation}
Approaching 2D-QSI criticality matching requires $\Delta \left(
\xi _{c0}/\Lambda _{ab}^{2}\right) /\left( \xi _{c0}/\Lambda
_{ab}^{2}\right) \rightarrow \Delta d_{s}/d_{s}$. For the oxygen
isotope effect ($^{16}$O vs. $^{18}$O) on a physical quantity $X$
\ the relative isotope shift is defined as $\Delta
X/X=(^{18}X-^{16}X)/^{18}X$. In this case the effect has been
traced back to a change of the critical amplitude of the c-axis
correlation length upon oxygen isotope exchange. Another
implication is that the absence of a substantial isotope effect on
the transition temperature, e.g. close to optimum doping, is not
of particular significance. In this case what is left is the
effect on the in-plane penetration depth.

\begin{figure}[tbp]
\centering
\includegraphics[totalheight=6cm]{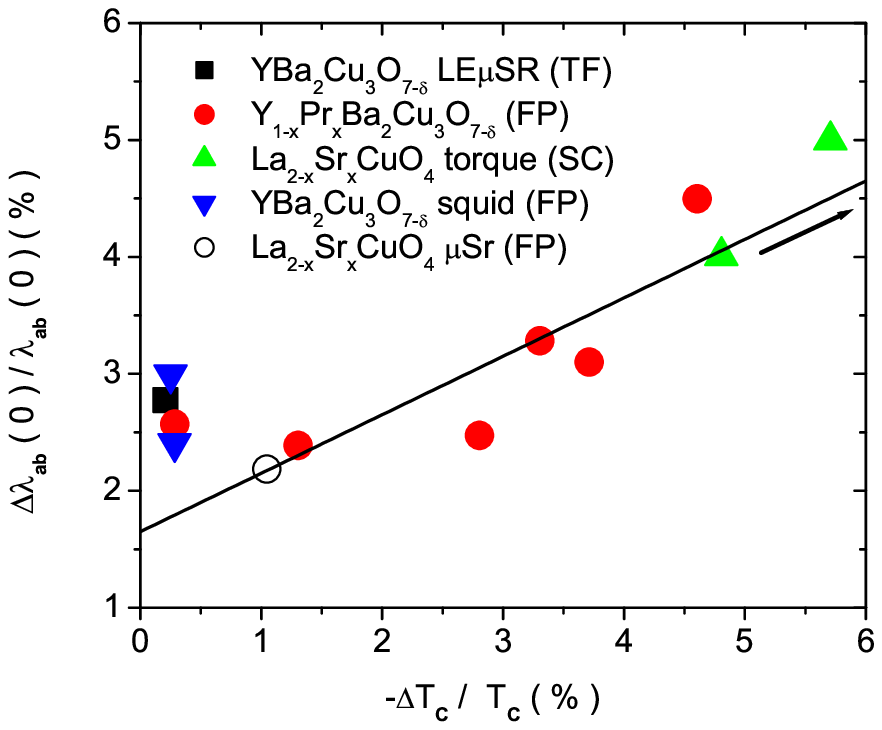}
\caption{Data for the oxygen isotope effect in underdoped
La$_{2-x}$Sr$_{x}$CuO$_{4}$($\bigcirc $:
$x$=0.15\protect\cite{zhao1}, $\blacktriangle $: $x$=0.08, 0.086
\protect\cite{hofer214}),
Y$_{1-x}$Pr$_{x}$Ba$_{2}$Cu$_{3}$O$_{7-\delta }$ ($\bullet $:
$x$=0, 0.2, 0.3, 0.4\protect\cite{zhao1,khasanov123pr,rksite}) and
YBa$_{2}$Cu$_{3} $O$_{7-\delta }$ ($\blacktriangledown $
\protect\cite{zhao1}, $\blacksquare $\protect\cite {khasanov123f})
\ in terms of $\Delta \left( \lambda _{ab}\left( 0\right) \right)
/$ $\lambda _{ab}\left( 0\right) $ versus -$\Delta T_{c}/T_{c}$.
In the direction of the arrow the solid line indicates the flow to
2D-QSI criticality and provides with Eq. (\ref{eq22}) an estimate
for the oxygen isotope effect on $d_{s}$, namely $\Delta
d_{s}/d_{s}=3.3(4)\%$.} \label{fig11}
\end{figure}

In Fig. \ref{fig11} we show the data for the oxygen isotope effect
in La$_{2-x}$Sr$_{x}$CuO$_{4}$\cite{hofer214,zhao1},
Y$_{1-x}$Pr$_{x}$Ba$_{2}$Cu$_{3}$O$_{7-\delta
}$\cite{zhao1,khasanov123pr,rksite} and
YBa$_{2}$Cu$_{3}$O$_{7-\delta }$\cite{zhao1,khasanov123f},
extending from the underdoped to the optimally doped regime, in
terms of $\Delta \left( \lambda _{ab}\left( 0\right) \right) /$
$\lambda _{ab}\left( 0\right) $ versus $\Delta T_{c}/T_{c}$. It is
evident that there is a correlation between the isotope effect on
$T_{c}$ and $\lambda _{ab}\left( 0\right) $. Indeed, approaching
the 2D-QSI transition, as marked by the arrow, the data tends to
fall on the straight line, which is Eq. (\ref{eq22}). This yields
for the isotope effect on $d_{s}$ the estimate $\Delta
d_{s}/d_{s}=3.3(4)\%$. Approaching optimum doping, this
contribution renders the isotope effect on $T_{c}$ considerably
smaller than that on $\lambda _{ab}\left( 0\right) $. In fact,
even in nearly optimally doped YBa$_{2}$Cu$_{3}$O$_{7-\delta }$,
where $\Delta T_{c}/T_{c}=-0.26(5)\%$, a substantial isotope
effect on the in-plane penetration depth, $\Delta \lambda
_{ab}\left( 0\right) /\lambda _{ab}\left( 0\right) =-2.8(1.0)\%$,
has been established by direct observation, using the novel
low-energy muon-spin rotation technique\cite{khasanov123f}.

\begin{figure}[tbp]
\centering
\includegraphics[totalheight=6cm]{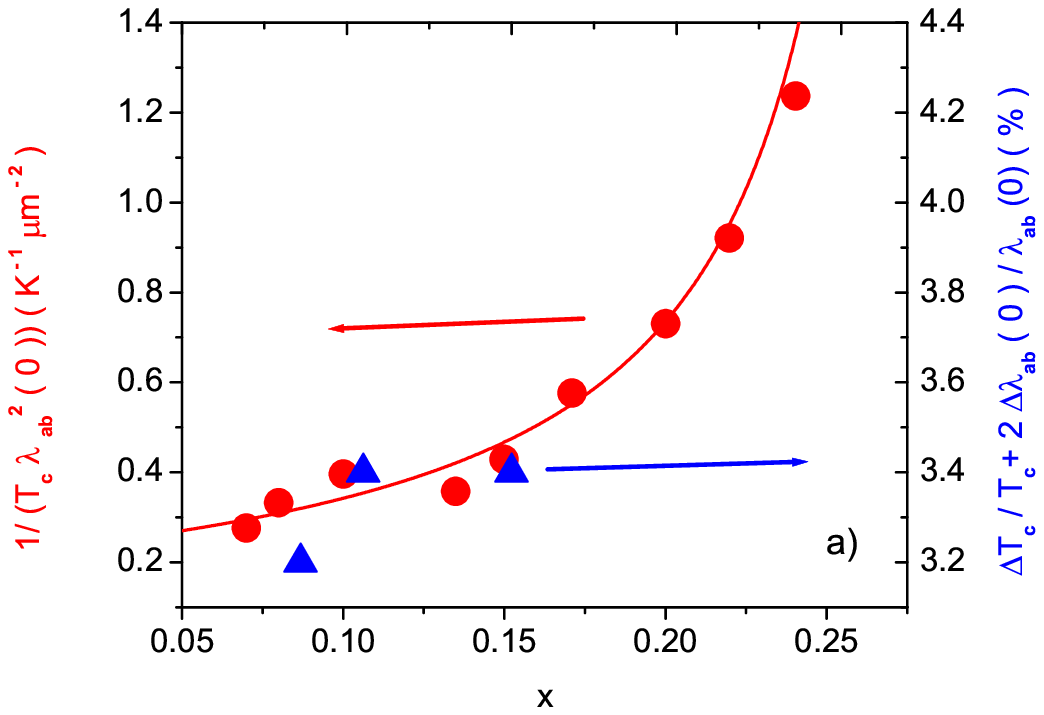}
\includegraphics[totalheight=6cm]{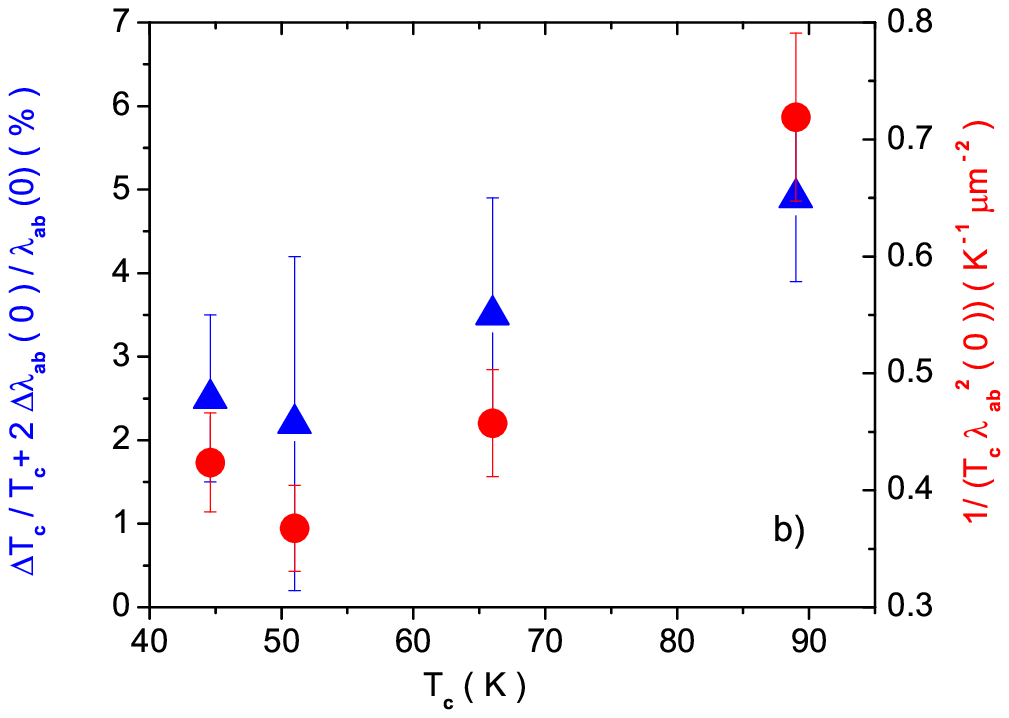}
\caption{a) Data for the oxygen isotope effect in
La$_{2-x}$Sr$_{x}$CuO$_{4}$($\blacktriangle :$ x= 0.086
\protect\cite{hofer214}, x=0.105\protect\cite{zhao1},
0.15\protect\cite {khasanov214} in terms of $\Delta
T_{c}/T_{c}+2\Delta \left( \lambda _{ab}\left( 0\right) \right) /$
$\lambda _{ab}\left( 0\right) $ \textit{vs.} $x$. For comparison
we included the experimental data ($\bullet $) for $1/\xi
_{c0}\propto 1/\left( T_{c}\lambda _{ab}^{2}\left( 0\right)
\right) $ \textit{vs.} $x$ and the solid curve is taken from Eq.
(\ref{eq12}). b) $\Delta T_{c}/T_{c}+2\Delta \left( \lambda
_{ab}\left( 0\right) \right) /$ $\lambda _{ab}\left( 0\right) $
\textit{vs.} $T_{c}$ of
Y$_{1-x}$Pr$_{x}$Ba$_{2}$Cu$_{3}$O$_{7-\delta }$ ($\blacktriangle
$: x=0, 0.2, 0.3, 0.4)\protect\cite{tsrkhk,khasanovmsr}. The
experimental data ($\bullet $) for $1/\xi _{c0}\propto 1/\left(
T_{c}\lambda _{ab}^{2}\left( 0\right) \right) $\textit{vs.}
$T_{c}$ are included for comparison.} \label{fig12}
\end{figure}

Fig. \ref{fig11} also shows that away from 2D-QSI criticality the
experimental data does no longer fall onto the straight line.
Since in the underdoped regime $\Delta \left( \xi _{c0}/\Lambda
_{ab}^{2}\right) /\left( \xi _{c0}/\Lambda _{ab}^{2}\right) $
$\rightarrow \Delta d_{s}/d_{s}$ this behavior unfolds according
to Eq. (\ref{eq23}) the doping dependence of $\Delta \left( \xi
_{c0}/\Lambda _{ab}^{2}\right) /\left( \xi _{c0}/\Lambda
_{ab}^{2}\right) $. To disentangle the doping dependence of
$\Delta \left( \xi _{c0}/\Lambda _{ab}^{2}\right) $ and $\xi
_{c0}/\Lambda _{ab}^{2}$ we displayed in Fig. \ref{fig12}a $\Delta
T_{c}/T_{c}+2\Delta \left( \lambda _{ab}\left( 0\right) \right) /$
$\lambda _{ab}\left( 0\right) $ \textit{vs.} $x$ for the oxygen
isotope in La$_{2-x}$Sr$_{x}$CuO$_{4}$. For comparison we included
$1/\xi _{c0}\propto 1/\left( T_{c}\lambda _{ab}^{2}\left( 0\right)
\right) $\textit{vs.} $x$, indicating that $\Delta
T_{c}/T_{c}+2\Delta \left( \lambda _{ab}\left( 0\right) \right) /$
$\lambda _{ab}\left( 0\right) $ scales as $1/\xi _{c0}\propto
1/\left( T_{c}\lambda _{ab}^{2}\left( 0\right) \right) $and
accordingly $\Delta \xi _{c0}/\xi _{c0}$ as $\Delta \xi _{c0}/\xi
_{c0}\propto 1/\xi _{c0}$. Hence $\Delta \xi _{c0}$ is essentially
independent of the dopant concentration.

\begin{figure}[tbp]
\centering
\includegraphics[totalheight=6cm]{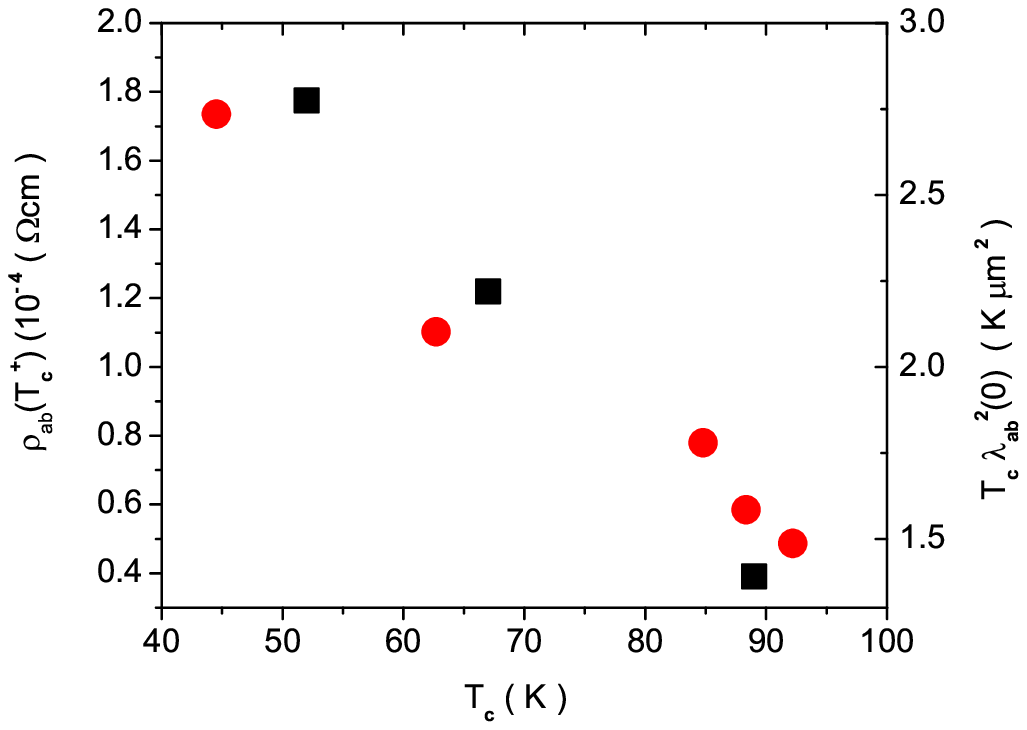}
\caption{$T_{c}\lambda _{ab}^{2}\left( 0\right) $ \textit{vs.}
$T_{c}$ for Y$_{1-x}$Pr$_{x}$Ba$_{2}$Cu$_{3}$O$_{7-\delta }$ taken
from \protect\cite{tsrkhk} and $\rho _{ab}\left( T_{c}^{+}\right)
$ \textit{vs.} $T_{c}$ taken from Maple \emph{et
al}.\protect\cite{maple} and Dalichaouch \emph{et al.}
\protect\cite{dali}.} \label{fig13}
\end{figure}

Clearer evidence for this scaling behavior of the isotope effect
on the amplitude of the c-axis correlation length emerges from
Fig. \ref{fig12}b showing $\Delta T_{c}/T_{c}+2\Delta \left(
\lambda _{ab}\left( 0\right) \right) /$ $\lambda _{ab}\left(
0\right) $ \textit{vs.} $T_{c}$ and $1/\xi _{c0}\propto 1/\left(
T_{c}\lambda _{ab}^{2}\left( 0\right) \right) $\textit{vs.}
$T_{c}$ for Y$_{1-x}$Pr$_{x}$Ba$_{2}$Cu$_{3}$O$_{7-\delta }$. This
opens a window onto the origin of the unconventional isotope
effects in the cuprates. Indeed, there is the evidence for a
nearly linear doping dependence of $\xi _{c0}$, namely $\xi
_{c0}\approx d_{s}-b\delta $ (see Eq.(\ref{eq12})) and Figs.
\ref{fig5}a, \ref{fig6} and \ref{fig9}a) and the limiting behavior
$\Delta \xi _{c0}\rightarrow \Delta d_{s}$ (Eq. (\ref{eq16}). To
substantiate this point we displayed in Fig. \ref{fig13} the
experimental data for
Y$_{1-x}$Pr$_{x}$Ba$_{2}$Cu$_{3}$O$_{7-\delta }$ in terms of
$T_{c}\lambda _{ab}^{2}\left( 0\right) $ \textit{vs.} $T_{c}$
\cite{tsrkhk} and $\rho _{ab}\left( T_{c}^{+}\right) $
\textit{vs.} $T_{c}$\cite {maple,dali}. Apparently, the scaling
relations (\ref{eq19}) and (\ref{eq20}), requiring $\xi
_{c0}\propto T_{c}\lambda _{ab}^{2}\left( 0\right) $ $\propto $
$\rho _{ab}\left( T_{c}^{+}\right) $ are well confirmed.

\begin{figure}[tbp]
\centering
\includegraphics[totalheight=6cm]{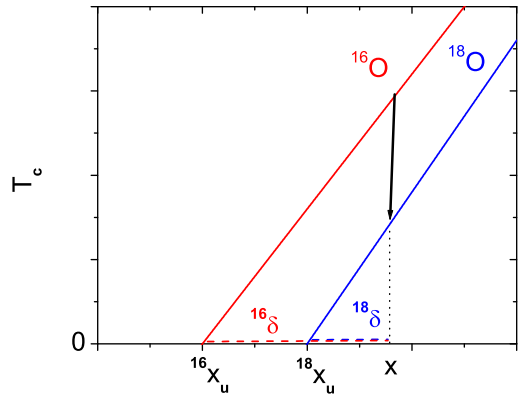}
\caption{Schematic sketch of the oxygen isotope effect on the
phase diagram of La$_{2-x}$Sr$_{x}$CuO$_{4}$ close to the
underdoped limit where the 2D-QSI transition occurs. The solid
lines are the phase transition lines $T_{c}\left( x\right) $.
Since the magnitude of $\Delta T_{c}=^{16}T_{c}-^{18}T_{c}$
increases with reduced $x$ the underdoped limit shifts from
$^{16}x_{u}$ to $^{18}x_{u}$. The arrow at fixed $x$ indicates the
reduction of $T_{c}$ upon complete oxygen isotope exchange
($^{16}$O $\rightarrow $ $^{18}$O) and the dashed lines the mobile
carrier concentration $^{16,18}\delta =x-^{16,18}x_{u}$.}
\label{fig14}
\end{figure}

Together with the weak doping dependence of $\Delta \xi _{c0}$
resulting from the evidence for $\Delta \xi _{c0}/\xi _{c0}\propto
1/\xi _{c0}$(Fig. \ref{fig12}b) these constraints imply that
\begin{equation}
\Delta \xi _{c0}\approx \Delta d_{s}-\Delta b\delta ^{n}-bn\delta
^{n-1}\Delta \delta =\Delta d_{s}+bn\delta ^{n-1}\Delta
x_{u},\text{ }n\gtrsim 1,  \label{eq25}
\end{equation}
where $\delta =x-x_{u}$ is the concentration of the mobile charge
carriers. To appreciate the implications it is instructive to
consider the phase diagram of La$_{2-x}$Sr$_{x}$CuO$_{4}$
displayed in Fig.\ref{fig14}.

Given the experimental fact that $x$ and the lattice constants do
not change upon complete oxygen isotope exchange ($^{16}$O
$\rightarrow $ $^{18}$O)\cite {conder,raffa}, while $T_{c}$ is
lowered and $\Delta T_{c}$ increases by approaching the underdoped
limit, it follows that this limit shifts from $^{16}x_{u}$ to
$^{18}x_{u}$. Hence, for fixed $x$ La$_{2-x}$Sr$_{x}$CuO$_{4}$
becomes in the interval $^{16}x_{u}<x<$ $^{18}x_{u}$ an insulator
and the shift of the underdoped limit from $^{16}x_{u}$ to
$^{18}x_{u}$ leads to a reduction of the concentration $\delta $
of the mobile charge carriers, as indicated in Fig. \ref{fig14}.
The lesson then is, the unconventional isotope effects in the
cuprates stem from the shifts of $d_{s}$ and the underdoped limit
$x_{u}$. Since the volume of the unit cell is preserved this
changes imply local lattice distortions. Consequently, the isotope
shifts of $T_{c}$ and $\lambda _{ab}\left( 0\right) $ are not
related by scaling relations imposed by 3D-XY universality only,
but reveal the existence and relevance of the coupling between the
superfluid and volume preserving local lattice distortions.

This lesson appears to contradict interpretations based on the
London formula \cite{varenna}
\begin{equation}
\frac{1}{\lambda _{ii}^{2}\left( 0\right) }=\frac{4\pi
n_{s}e^{2}}{m_{ii}^{\ast }c^{2}},  \label{eq25a}
\end{equation}
where $n_{s}$ is the number density of the superfluid and
$m_{ii}^{\ast }$ the effective mass of the pairs in direction $i$.
There is the experimental evidence that the lattice constants do
not change upon complete oxygen isotope exchange ($^{16}$O
$\rightarrow $ $^{18}$O)\cite {conder,raffa}. Furthermore, recent
nuclear quadrupole resonance (NQR) studies of $^{16}$O and
$^{18}$O substituted optimally doped YBa$_{2}$Cu$_{3}
$O$_{7-\delta \text{ }}$powder samples revealed that the change of
the total charge density caused by the isotope exchange is less
than 10$^{-3}$\cite {khasanov123f}. These experimental facts have
then be taken as evidence for a negligibly small isotope effect on
$n_{s}$\cite{khasanov123f}. However, in general the London
relation is just a way of parameterizing experimental results,
with no discernible connection to the carrier concentration or,
e.g. the band mass. Indeed, even in conventional superconductors
$1/\lambda _{ii}^{2}\left( 0\right) $ is determined by normal
state properties, namely the integral of the Fermi velocity over
the Fermi surface. In the special case of spherical and
ellipsoidal Fermi surfaces one recovers Eq. (\ref {eq25a}) in the
form, $1/\lambda _{ii}^{2}\left( T=0\right) =4\pi ne^{2}/\left(
m_{ii}^{\ast }c^{2}\right) $, where $n$ is the number density of
the electrons in the normal state and $m_{ii}^{\ast }$ the band
mass \cite {tinkham}. In any case, since the Hall effect is
related to the concentration of mobile charge
carriers\cite{gariglio} the reduction of $\delta $ upon isotope
exchange should lead to an isotope effect on the Hall constant
$R_{H}$. Within the framework of the model proposed by Stojkovic
and Pines\cite{stoikovic}, where both the resistivity and the Hall
constant are inversely proportional to $\delta $, the relative
isotope shifts are then related by
\begin{equation}
\frac{\Delta R_{H}}{R_{H}}=\frac{\Delta \rho }{\rho
}=-\frac{\Delta \delta }{\delta }.  \label{eq25b}
\end{equation}
Although qualitative evidence for the oxygen isotope effect on the
resistivity emerges from the measurements on
Y$_{1-x}$Pr$_{x}$Ba$_{2}$Cu$_{3} $O$_{7-\delta }$\cite{franck} and
on Pr-, Ca-, and Zn-substituted YBa$_{2}$Cu$_{3}$O$_{7-\delta }$
\cite{sorensen}, this relationship awaits to be tested
quantitatively.

Further evidence for this coupling emerges from the combined
isotope and finite size effects. Recently, it has been shown that
the notorious rounding of the superconductor to normal state
transition is fully consistent with a finite size effect,
revealing that bulk cuprate superconductors break into nearly
homogeneous superconducting grains of rather unique extent\cite
{book,bled,tsdc,tsrkhk}. A characteristic feature of this finite
size effect on the temperature dependence of the in-plane
penetration depth $\lambda _{ab}$ is the occurrence of an
inflection point giving rise to an extremum in $d\left( \lambda
_{ab}^{2}\left( T=0\right) /\lambda _{ab}^{2}\left( T\right)
\right) /dT$ at $T_{p_{c}}$. Here $\lambda _{ab}^{2}\left(
T_{p_{c}}\right) $, $T_{p_{c}}$ and the length $L_{c}$ of the
grains along the c-axis are related by\cite{bled,tsdc,tsrkhk}
\begin{equation}
\frac{1}{\lambda _{ab}^{2}\left( T_{p}\right) }=\frac{16\pi
^{3}k_{B}T_{p_{c}}}{\Phi _{0}^{2}L_{c}}.  \label{eq26}
\end{equation}
Recently, the effect of oxygen isotope exchange on $L_{c}$ has
been studied in Y$_{1-x}$Pr$_{x}$Ba$_{2}$Cu$_{3}$O$_{7-\delta
}$\cite{tsrkhk}. Note that the relative shifts of $\lambda
_{ab}^{2}\left( T_{p_{c}}\right) $, $T_{p_{c}}$ and $L_{c}$ are
not independent but according to Eq. (\ref{eq26}) related by
\begin{equation}
\frac{\Delta L_{c}}{L_{c}}=\frac{\Delta
T_{p_{c}}}{T_{p_{c}}}+\frac{\Delta \lambda _{ab}^{2}\left(
T_{p_{c}}\right) }{\lambda _{ab}^{2}\left( T_{p_{c}}\right)
}=\frac{\Delta T_{p_{c}}}{T_{p_{c}}}+2\frac{\Delta \lambda
_{ab}\left( T_{p_{c}}\right) }{\lambda _{ab}\left(
T_{p_{c}}\right) }, \label{eq27}
\end{equation}
which is just the finite size scaling counterpart of
Eq.(\ref{eq23}). Some estimates resulting from the finite size
scaling analysis are listed in Table I. Several observations
emerge. First, the spatial extent of the homogeneous domains
$L_{c}$ changes upon isotope exchange ($^{16}$O, $^{18}$O). To
appreciate the implications of this observation, we note again
that for fixed Pr concentration the lattice parameters remain
essentially unaffected \cite{conder,raffa}. Accordingly, an
electronic mechanism, without coupling to local lattice
distortions implies $\Delta L_{c}=0$. On the contrary, a
significant change of $L_{c}$ upon oxygen exchange requires local
lattice distortions involving the oxygen lattice degrees of
freedom and implies with Eq. (\ref{eq27}) a coupling between these
distortions and the superfluid, probed by $\lambda _{ab}\left(
T_{p_{c}}\right) $. Second, since the relative shift of
$T_{p_{c}}$ is very small the change of $L_{c}$ is essentially due
to the superfluid, probed by $\lambda _{ab}\left( T_{p_{c}}\right)
$. Third, $L_{c}$ increases systematically with reduced
$T_{p_{c}}$.
\begin{center}
\begin{tabular}{|c|c|c|c|}
\hline x & 0 & 0.2 & 0.3 \\ \hline $\Delta L_{c}/L_{c}$ & 0.12(5)
& 0.13(6) & 0.16(5) \\ \hline $\Delta T_{p_{c}}/T_{p_{c}}$ &
-0.000(2) & -0.015(3) & -0.021(5) \\ \hline $\Delta \lambda
_{ab}^{2}\left( T_{p_{c}}\right) /\lambda _{ab}^{2}\left(
T_{p_{c}}\right) $ & 0.11(5) & 0.15(6) & 0.15(5) \\ \hline
$^{16}T_{p_{c}}$(K) & 89.0(1) & 67.0(1) & 52.1(1) \\ \hline
$^{16}L_{c}$(\AA ) & 9.7(4) & 14.2(7) & 19.5(8) \\ \hline
$^{16}\lambda _{ab}\left( 0\right) $(\AA ) & 1250(10) & 1820(20) &
2310(30)
\\ \hline
\end{tabular}
\end{center}

Table I: Finite size estimates for the relative changes of
$L_{c}$, $T_{p_{c}}$ and $\lambda _{ab}^{2}\left( T_{p_{c}}\right)
$ upon oxygen isotope exchange for
Y$_{1-x}$Pr$_{x}$Ba$_{2}$Cu$_{3}$O$_{7-\delta }$\cite {tsrkhk}.
\bigskip
We have seen that the unconventional isotope shifts of $T_{c}$ ,
$\lambda _{ab}\left( 0\right) $ and $L_{c}$ are not related by
scaling relations imposed by 3D-XY universality only, but reveal
the existence and relevance of the coupling between the superfluid
and volume preserving local lattice distortions. Their response to
isotope exchange shifts $d_{s}$, $L_{c}$ and the underdoped limit
$x_{u}$, accompanied by a reduction of the concentration $\delta $
of mobile carriers. These observations contradict the majority
opinion on the mechanism of superconductivity in the cuprates that
it occurs via a purely electronic mechanism involving spin
excitations, and the lattice degrees of freedom are irrelevant.

\begin{figure}[tbp]
\centering
\includegraphics[totalheight=6cm]{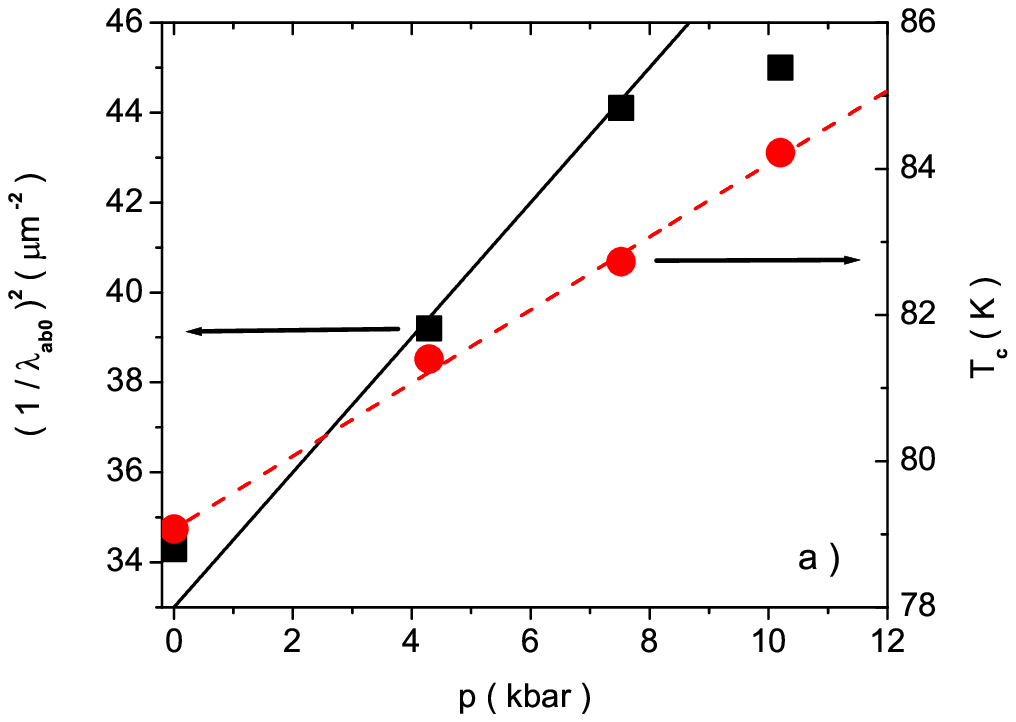}
\includegraphics[totalheight=6cm]{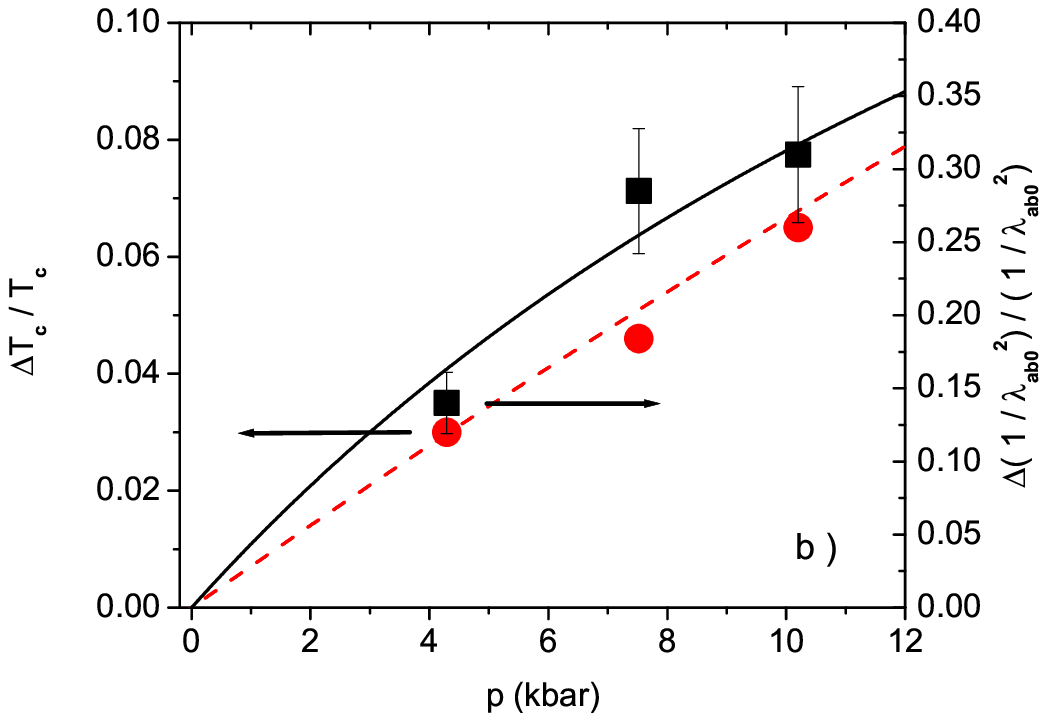}
\caption{a) Pressure dependence of $T_{c}$ and $1/$ $\lambda
_{ab0}^{2}$ in YBa$_{2}$Cu$_{4}$O$_{8}$ taken from Khasanov
\emph{et al.}\protect\cite{khasanovp124}. The dashed and solid
lines are given by Eq. (\ref{eq28}). b) Pressure dependence of
$\Delta T_{c}/T_{c}$ and $\Delta \left( 1/\lambda
_{ab0}^{2}\right) /\left( 1/\lambda _{ab0}^{2}\right) =-2\Delta
\lambda _{ab0}/\lambda _{ab0}$. The solid and dashed lines follow
from Eq. (\ref {eq28}).} \label{fig15}
\end{figure}

As aforementioned the exact 3D-XY scaling relation (\ref{eq17})
and its approximate counterpart (\ref{eq18}) should also apply
when pressure is applied. As does the isotope effect, there is a
generic decrease of $T_{c}$ with pressure in simple-metal
superconductors (Sn, Al, In, \textit{etc}.). In cuprate
superconductors the situation is considerably more complex because
changes in both the local lattice distortions and lattice
constants occur, and there are pressure-induced relaxation
phenomena. In YBa$_{2}$Cu$_{3}$O$_{7-_{\delta
}}$\cite{fietz,sadewasser}, Tl$_{2}$Ba$_{2}$CuO$_{6+\delta
}$\cite{sieburger,klehe,looney}, and other cuprates\cite
{schilling} the initial dependence of $T_{c}$ on pressure often
depends markedly on the pressure-temperature history of the
sample. The relaxation phenomena responsible for this behavior are
believed to originate from pressure-induced ordering of mobile
oxygen defects in the lattice and the value of $T_{c}$ appears to
be a sensitive function of both the concentration and the
arrangement of these defects. To avoid these difficulties we
concentrate on YBa$_{2}$Cu$_{4}$O$_{8}$, exhibiting a spectacular
increase of $T_{c}$ under pressure by nearly
$30$K\cite{griessen,schilling2}. Furthermore, the pressure
dependence of the critical amplitude $\lambda _{ab0}$ have been
studied as well\cite{khasanovp124}. The data displayed in Fig.
\ref{fig15}a show that the pressure dependencies of $T_{c}$ and
$1/$ $\lambda _{ab0}^{2}$ are initially nearly linear and
positive, as indicated by the solid and dashed lines given by
\begin{equation}
T_{c}\left( p\right) =79.07+0.5p,\text{ }1/\lambda
_{ab0}^{2}\left( p\right) =33+1.5p.  \label{eq28}
\end{equation}
As a result both $\Delta T_{c}/T_{c}$ and $\Delta \left( 1/\lambda
_{ab0}^{2}\right) /\left( 1/\lambda _{ab0}^{2}\right) =-2\Delta
\lambda _{ab0}/\lambda _{ab0}$ are positive and increase with
pressure, as shown in Fig. \ref{fig15}b. This differs from the
oxygen isotope effect where these quantities are negative (see
Fig. \ref{fig11}). Nevertheless, the universal relation
(\ref{eq17}) implies that these pressure induced changes are not
independent but related by
\begin{equation}
\frac{\Delta T_{c}}{T_{c}}+2\frac{\Delta \lambda _{ab0}}{\lambda
_{ab0}}=\frac{\Delta \xi _{c0}}{\xi _{c0}}.  \label{eq29}
\end{equation}

\begin{figure}[tbp]
\centering
\includegraphics[totalheight=6cm]{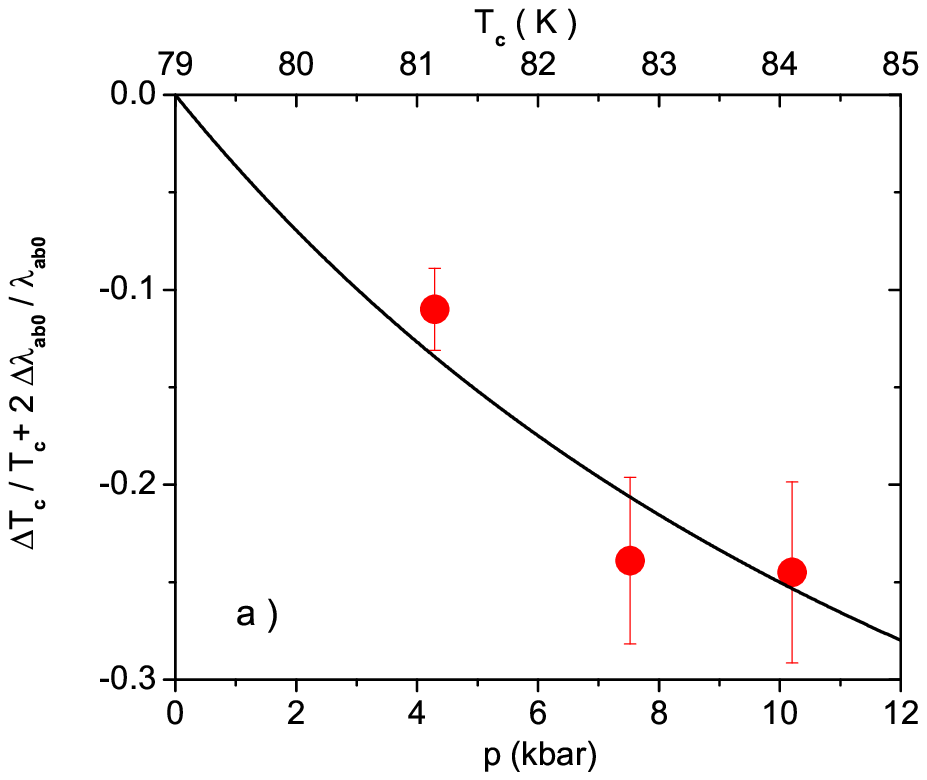}
\includegraphics[totalheight=6cm]{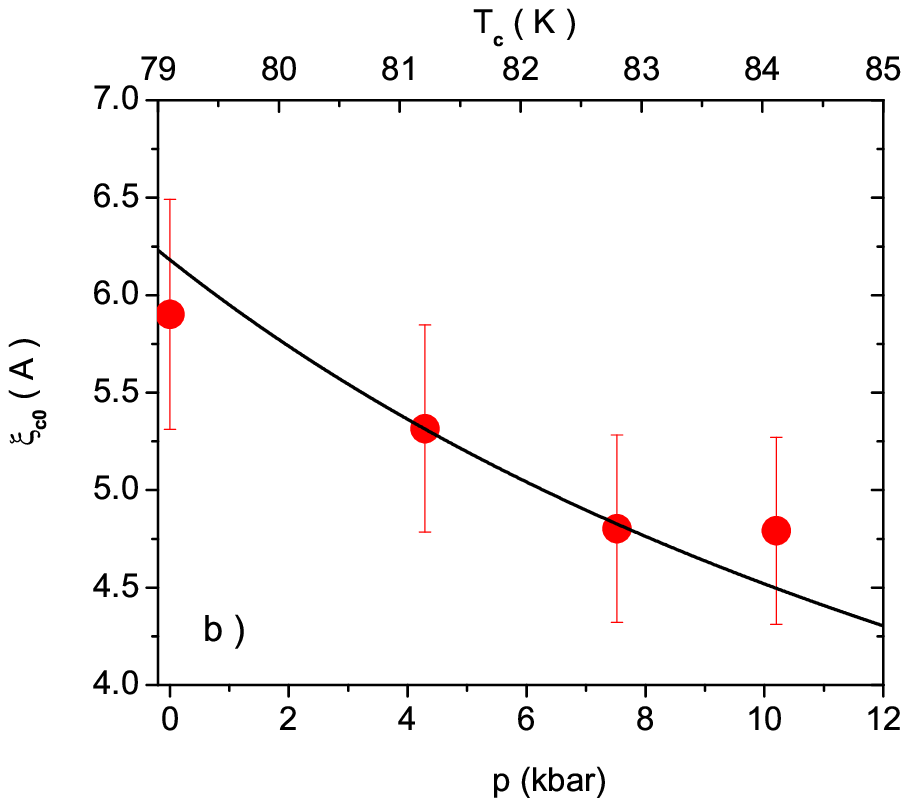}
\caption{a) $\Delta T_{c}/T_{c}+2\Delta \lambda _{ab0}/\lambda
_{ab0}$ \textit{vs.} $p$ and $T_{c}$ for YBa$_{2}$Cu$_{4}$O$_{8}$
derived from the data of Khasanov \emph{et
al.}\protect\cite{khasanovp124}. The solid line follows from Eq.
(\ref{eq28}); b) $\xi _{c0}$ \textit{vs.} $p$ and $T_{c}$ for
YBa$_{2}$Cu$_{4}$O$_{8}$ derived from the data of Khasanov
\emph{et al.}\protect\cite {khasanovp124} with the aid of Eq.
(\ref{eq9}). The solid line follows from Eqs. (\ref{eq17}) and
(\ref{eq28}).} \label{fig16}
\end{figure}

In Fig. \ref{fig16}a we displayed the experimental estimates for
$\Delta T_{c}/T_{c}+2\Delta \lambda _{ab0}/\lambda _{ab0}$
\textit{vs.} $p$ and $T_{c}$ for YBa$_{2}$Cu$_{4}$O$_{8}$ derived
from the data of Khasanov \emph{et al.}\cite{khasanovp124}. For
comparison we included the behavior resulting from Eq.
(\ref{eq28}). To interpret the data we invoke Eq. (\ref {eq9}) to
derive from $T_{c}\left( p\right) $ and $\lambda _{ab0}\left(
p\right) $ the pressure dependence of $\xi _{c0}\left( p\right) $.
A glance to Fig. \ref{fig16}b shows that $\xi _{c0}$ decreases
with increasing pressure and transition temperature. Noting that
YBa$_{2}$Cu$_{4}$O$_{8}$ falls at zero pressure into the slightly
underdoped regime this behavior can be compared with the $T_{c}$
dependence of $\xi _{c0}$ in underdoped
La$_{2-x}$Sr$_{x}$CuO$_{4}$ displayed in Fig. \ref{fig5}a. In this
regime the rise of $T_{c}$ implies a reduction of $\xi _{c0}$.
Considering the behavior in the $\left( \xi _{c0},\text{
}T_{c}\right) $- plane the hydrostatic pressure effect may then
also be viewed as an inverse oxygen isotope effect ($^{18}$O
$\rightarrow $ $^{16}$O), a pressure induced crossover to less
anisotropic 3D behavior, or a crossover to optimum doping. This
crossover is also related to the initial decrease of the
$a$-Lattice constant under applied hydrostatic
pressure\cite{tang}. Thus, we arrive at the following tentative
scenario. If the layer is originally underdoped hydrostatic
pressure enhances $T_{c}$ and reduces both $\lambda _{ab0}$ and
$\xi _{c0}$; if it is overdoped, it decreases $T_{c}$ increases
$\lambda _{ab0}$ and reduces $\xi _{c0}$. This behavior can be
anticipated from from Figs. \ref {fig5}a, \ref{fig15} and
\ref{fig16} and uncovers again 3D-XY scaling and 2D to 3D
crossover to be at work.

\begin{figure}[tbp]
\centering
\includegraphics[totalheight=6cm]{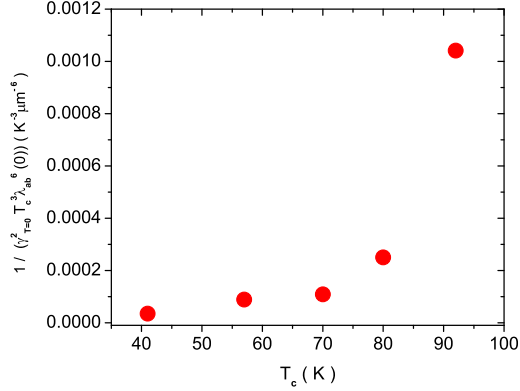}
\caption{$1/\left( \gamma _{T=0}^{2}T_{c}^{3}\lambda
_{ab}^{6}\left( 0\right) \right) $ \textit{vs.} $T_{c}$ for
YBa$_{2}$Cu$_{3}$O$_{7-\delta }$ derived from the data of Trunin
and Nefyodov\protect\cite{trunin}.} \label{fig17}
\end{figure}

So far we concentrated on scaling relations emerging from the
universal amplitude combination (\ref{eq17}). There is a multitude
of other combinations including the relation between
magnetization, applied magnetic field and $T_{c}$\cite{book,hub},
as well as the universal relation between the critical amplitude
$A^{+}$of the specific heat singularity and the correlation volume
$V_{c}^{+}$\cite{book,peliasetto}
\begin{equation}
\ A^{+}V_{c}^{+}=\left( R^{+}\right) ^{3},V_{c}^{+}=\xi _{a0}\xi
_{b0}\xi _{c0}.  \label{eq30}
\end{equation}
$\alpha $ is the critical exponent and $A^{\pm }$ the critical
amplitude of the specific heat singularity, $c=\left( A^{\pm
}/\alpha \right) \left| t\right| ^{-\alpha }+B^{\pm }$, where $\pm
=sgn(T/T_{c}-1)$. 3D-XY universality implies\cite {peliasetto}
\begin{equation}
\frac{A^{+}}{A^{-}}=1.07,\ \text{ }R^{+}=0.361,\text{ }\alpha
=2-3\nu =-0.013,\ \nu =0.671.  \label{eq31}
\end{equation}
Although the singular part of the specific heat is small compared
to the phonon contribution and inhomogeneities set limits in
exploring the critical regime\cite{book,bled} the measurements on
nearly optimally doped YBa$_{2}$Cu$_{3}$O$_{7-\delta
}$\cite{book,pasler} and La$_{2-x}$Sr$_{x}$CuO$_{4}$\cite{nyhusa}
point clearly to 3D-XY critical behavior. Experimentally it is
well established that the specific heat anomaly and with that the
critical amplitude $A^{+}$ decreases dramatically by approaching
the underdoped limit (2D-QSI
criticality)\cite{ghiron,loram,loram2}. Since $V_{c}^{+}=\xi
_{a0}\xi _{b0}\xi _{c0}\approx \gamma _{T_{c}}^{2}\xi _{c0}^{3}$
and $\xi _{c0}\rightarrow d_{s}$ ( see Figs. \ref{fig4},
\ref{fig5}a and \ref{fig7}b) while  $\gamma
_{T_{c}}^{2}\rightarrow \infty $ (see Eq. (\ref{eq2})) in the
underdoped limit, this behavior is a consequence of the
3D-2D-crossover. Furthermore, combining the universal 3D-XY
amplitude combinations (\ref{eq17}) and (\ref{eq30}) we obtain the
universal relation
\begin{equation}
A^{+}=\frac{\left( R^{+}\right) ^{3}}{\left( \gamma
_{T_{c}}\right) ^{2}\xi _{c0}^{3}}=\left( \frac{\Phi
_{0}^{2}f}{16\pi ^{3}k_{B}}\right) ^{3}\frac{1}{\gamma
_{T_{c}}^{2}\lambda _{ab0}^{6}T_{c}^{3}},  \label{eq32}
\end{equation}
relating the critical amplitudes of specific heat, anisotropy,
in-plane penetration depth and anisotropy. Consequently, the
pressure and isotope effects on these quantities are not
independent but related by\cite{tsiso}
\begin{equation}
\frac{\Delta A^{+}}{A^{+}}=-2\frac{\Delta \gamma _{T_{c}}}{\gamma
_{T_{c}}}-6\frac{\Delta \lambda _{ab0}}{\lambda
_{ab0}}-3\frac{\Delta T_{c}}{T_{c}}. \label{eq33}
\end{equation}
Although these universal 3D-XY relations await to be tested
experimentally they reveal again that as long as uncharged
fluctuations dominate these effects do not change $T_{c}$ only but
uncover the coupling between energy- and superfluid fluctuations.
To illustrate the dramatic decrease of the specific heat anomaly
in the underdoped regime we plotted in Fig. \ref{fig17} $1/\left(
\gamma _{T=0}^{2}T_{c}^{3}\lambda _{ab}^{6}\left( 0\right) \right)
$ \textit{vs.} $T_{c}$ for YBa$_{2}$Cu$_{3}$O$_{7-\delta }$
derived from the data of Trunin and Nefyodov\cite{trunin}. In the
heavily underdoped regime where $T_{c}\propto 1/\lambda
_{ab}^{2}\left( 0\right) $ ( see Eq. (\ref{eq5})) the magnitude of
\ $A^{+}$ is then manly controlled by the anisotropy $\gamma $
which diverges at 2D-QSI criticality (see Eq. (\ref{eq2}) and
Figs. \ref{fig1}b and \ref{fig2})

Another example is the universal relation between magnetization
$m$, applied magnetic field $H$ and $T_{c}$\cite{hoferdis,book},
\begin{equation}
m_{c}=\frac{Q_{3}C_{3}k_{B}}{\Phi _{0}^{3/2}}H_{c}T_{c}\gamma
_{T_{c}},\text{ }m_{ab}=\frac{Q_{3}C_{3}k_{B}}{\Phi
_{0}^{3/2}}H_{ab}\frac{T_{c}}{\gamma _{T_{c}}},  \label{eq34}
\end{equation}
for fields applied parallel $\left( H_{c}\right) $ or
perpendicular $\left( H_{ab}\right) $ to the c-axis. $Q_{3}C_{3}$
is a universal constant\cite {hoferdis,book}. Hence, the isotope
and pressure effects on these quantities are not independent but
related by
\begin{equation}
\frac{\Delta m_{c}}{m_{c}}=\frac{\Delta T_{c}}{T_{c}}+\frac{\Delta
\gamma _{T_{c}}}{\gamma _{T_{c}}},\text{ }\frac{\Delta
m_{ab}}{m_{ab}}=\frac{\Delta T_{c}}{T_{c}}-\frac{\Delta \gamma
_{T_{c}}}{\gamma _{T_{c}}}.  \label{eq35}
\end{equation}
Although the isotope effect on the magnetization is well
established\cite {hofer214,khasanov123pr,franck,sorensen} these
relationships await to be explored.

In summary, by taking experimental data for $T_{c}$, $\lambda
_{ab,c}\left( 0\right) $, $\sigma _{ab,c}^{dc}\left(
T_{c}^{+}\right) =1/\rho _{ab,c}^{dc}\left( T_{c}^{+}\right) $,
$\gamma =\lambda _{c}/\lambda _{ab}$ of a variety of cuprates
encompassing the underdoped and overdoped regimes, we have shown
that these quantities are not independent but related by scaling
relations reminiscent to anisotropic systems falling into the
3D-XY universality class. They differ from the exact scaling
relation by the ratio between the critical amplitudes of the
penetration depths and their zero temperature counterparts.
However, the remarkable agreement revealed that the doping and
family dependence of this ratio is rather weak. As the
unconventional isotope effects are concerned, the rather sparse
experimental data turned out to be consistent with the 3D-XY
scaling relations, connecting the isotope effects on $T_{c}$,
$\lambda _{ab}\left( 0\right) $ and the c-axis correlation length
$\xi _{c}$. Guided by the doping dependence of $\xi _{c}$ we have
shown that the oxygen isotope effects are associated with a change
of $d_{s}$, a shift of the underdoped limit and a reduction of the
concentration of mobile carriers. Since $\xi _{c}$ scales as $\xi
_{c}\propto \rho _{ab}\left( T_{c}^{+}\right) =1/\sigma
_{ab}\left( T_{c}^{+}\right) $ we obtained the approximate 3D-XY
scaling relation (\ref{eq4}). It connects the relative isotope
shifts of measurable properties and awaits together with Eq.
(\ref{eq25b}), relating the shifts of Hall constant, resistivity
and mobile carrier concentration upon oxygen isotope exchange, to
be tested quantitatively. Qualitative evidence for the oxygen
isotope effect on the resistivity emerges from the measurements on
Y$_{1-x}$Pr$_{x}$Ba$_{2}$Cu$_{3}$O$_{7-\delta }$\cite{franck} and
on Pr-, Ca-, and Zn-substituted YBa$_{2}$Cu$_{3}$O$_{7-\delta }$
\cite{sorensen}. Furthermore, we have seen that the combined
isotope and finite size effects open a door to probe the coupling
between local lattice distortions and superconductivity in terms
of the isotope shift of $L_{c}$, the spatial extent of the
homogeneous superconducting grains along the c-axis. Noting that
$\Delta L_{c}/L_{c}$ is rather large (see Table I) this change
revealed the coupling between superfluidity and local lattice
distortion and this coupling is likely important in understanding
the pairing mechanism. Further evidence for 3D-XY scaling to be at
work emerged from the effects of hydrostatic pressure on $T_{c}$
and $\lambda _{ab}\left( 0\right) $, as well as from the doping
dependence of the specific heat singularity. Although the
currently available experimental data of the pressure and isotope
effects, as well as on the critical amplitudes are rather sparse,
we have shown that a multitude of empirical correlations between
$T_{c}$, magnetic penetration lengths, resistivity, conductivity,
specific heat, \textit{etc}.  are fully consistent with the
universal critical amplitude combinations for anisotropic systems
falling into the 3D-XY universality class and undergo a crossover
to a 2D quantum superconductor to insulator transition (2D-QSI).
Not unexpectedly we have shown that these correlations are
controlled by the doping or $T_{c}$ dependence of the c-axis
correlation length $\xi _{c}$ and the anisotropy $\gamma _{T}$.
Although much experimental work remains to be done, measuring the
quantities of interest on the same sample, the remarkable
consistency with 3D-XY scaling and the crossover to the 2D-QSI
quantum critical point single out 3D and anisotropic microscopic
models which incorporate local lattice distortions, fall in the
experimentally accessible regime into the 3D-XY universality
class, and incorporate the crossover to 2D-QSI criticality where
superconductivity disappears.

\acknowledgments The author is grateful to D. Di Castro, S. Kohout
and J. Roos for very useful comments and suggestions on the
subject matter and J. M. Triscone for pertinent advice concerning
the Hall effect.

\end{document}